\documentclass[prd,twocolumn, nofootinbib, superscriptaddress
]{revtex4-1}
\usepackage{longtable}
\usepackage{commath}
\usepackage{hyperref}

\usepackage{tabularx}
\usepackage{amsmath}
\usepackage{float}
\usepackage{datetime}
\usepackage{textpos}
\usepackage{booktabs}
\usepackage{graphicx}
\usepackage{mathrsfs}

\usepackage[usenames,dvipsnames]{color}

\usepackage[normalem]{ulem}

\newcommand{\apjs}{ApJS}

\newcommand{\mnras}{MNRAS}

\begin{document}
\title{%
 The Spatial Morphology of the Secondary Emission in the Galactic Center Gamma-Ray Excess 
 }
\author{Thomas Lacroix}
\affiliation{Institut d'Astrophysique de Paris, UMR 7095, CNRS, UPMC Universit\'{e} Paris 6, Sorbonne Universit\'{e}s, 98 bis boulevard Arago, 75014 Paris, France}
\author{Oscar Macias}
\affiliation{Center for Neutrino Physics, Department of Physics, Virginia Tech, Blacksburg, VA 24061, USA}
\affiliation{Instituto de F\'{i}sica, Universidad de Antioquia, Calle 70 No. 52-21, Medell\'{i}n, Colombia}
\author{Chris Gordon}
\affiliation{Department of Physics and Astronomy, Rutherford Building, University of Canterbury, Private Bag 4800, Christchurch 8140, New Zealand}
\author{Paolo Panci}
\affiliation{Institut d'Astrophysique de Paris, UMR 7095, CNRS, UPMC Universit\'{e} Paris 6, Sorbonne Universit\'{e}s, 98 bis boulevard Arago, 75014 Paris, France}
\author{C\'{e}line B\oe hm}
\affiliation{Institute for Particle Physics Phenomenology, Durham University, Durham, DH1 3LE, United Kingdom}
\affiliation{LAPTH, Universit\'{e} de Savoie, CNRS, BP 110, 74941 Annecy-Le-Vieux, France}
\author{Joseph Silk}
\affiliation{Institut d'Astrophysique de Paris, UMR 7095, CNRS, UPMC Universit\'{e} Paris 6, Sorbonne Universit\'{e}s, 98 bis boulevard Arago, 75014 Paris, France}
\affiliation{AIM-Paris-Saclay, CEA/DSM/IRFU, CNRS, Univ. Paris VII, F-91191 Gif-sur-Yvette, France}
\affiliation{The Johns Hopkins University, Department of Physics and Astronomy,
	3400 N. Charles Street, Baltimore, Maryland 21218, USA}
\affiliation{Beecroft Institute of Particle Astrophysics and Cosmology, Department of Physics,
	University of Oxford, Denys Wilkinson Building, 1 Keble Road, Oxford OX1 3RH, United Kingdom}

\begin{abstract}
Excess GeV gamma rays from the Galactic Center (GC) have been measured with the Fermi Large Area Telescope (LAT). The presence of the GC excess (GCE)  appears to be robust with respect to changes in the diffuse galactic background modeling. The three main proposals for the GCE are an unresolved population of millisecond pulsars (MSPs),  outbursts of cosmic rays from the GC region, and self-annihilating dark matter (DM). The injection of secondary electrons and positrons into the interstellar medium (ISM) by an unresolved population of MSPs or DM annihilations can lead to observable gamma-ray emission via inverse Compton scattering or bremsstrahlung. Here we investigate how to determine whether secondaries are important in a model for the GCE. We develop a method of testing model fit which accounts for the different spatial morphologies of the secondary emission. We examine several models which give secondary emission and illustrate a case where a broadband analysis is not sufficient to determine the need for secondary emission.
\end{abstract}

\maketitle

\section{Introduction}
\label{sec:introduction}

Several independent groups have reported evidence of extended spherically symmetric excess gamma-ray emission  above the diffuse galactic background (DGB) from the central few degrees around the Galactic Center (GC)~\citep{Goodenough2009gk,Vitale:2009hr,Hooper:2010mq,hooper,AbazajianKaplinghat2012,AbazajianKaplinghat2013,GordonMacias2013,FermiGalacticCenter2015}. The spectrum of this Galactic Center excess (GCE) peaks around 1-3 GeV and is harder than a pion bump. The GCE drops like $\sim \theta^{-2\gamma+1}$ where $\theta$ is the angle from the GC. This corresponds to a drop with radius $(r)$ from the GC  as $r^{-2\gamma}$ where $\gamma\sim 1.2$. The GCE has also been found to extend out as far as $\sim 10^\circ$ \cite{hooperslatyer2013,Daylan:2014} and its existence appears  to be robust with respect to systematic errors {  \cite{GordonMacias2013,MaciasGordon2014,Daylan:2014,Abazajian2014,
Zho2015,CaloreCholisWeniger2015,FermiGalacticCenter2015}}, although see Ref.~\cite{deBoer2015} for a counter-argument based on a spectral-only approach.  There is some debate over how much the spectrum and spatial morphology is affected by uncertainties in the DGB, with some authors arguing the effect could be quite large \cite{Gaggero2015,CarlsonLindenProfumo2015,FermiGalacticCenter2015,CarlsonLindenProfumo2016}. An additional independent ridge-like GeV excess which is correlated with the HESS TeV ridge~\cite{Aharonian:2006} has also been detected \cite{Hooper:2010mq,Yusef-Zadeh2013,MaciasGordon2014,Abazajian2014} and is thought to be due to cosmic rays interacting with molecular gas \cite{Yusef-Zadeh2013,MaciasGordon2014,YoastHullGallagherZweibel2014,MacGorCroProf2015}. 

Various alternative explanations of the GCE have been proposed. One possibility is a population of $\sim 10^3$ millisecond pulsars (MSPs) \cite{Abazajian:2010zy,AbazajianKaplinghat2012,Wharton2012,GordonMacias2013,GordonMacias2013erratum,Mirabal2013,MaciasGordon2014,YuanZhang2014,BrandtKocsis2015} or young pulsars \cite{OLeary2015}. However, there is some debate about whether the subsequent emission could successfully extend out as far as $\sim 10^\circ$ \cite{Hooper2013,CholisHooperLinden2015,PetrovicSerpicoZaharijas2015,Lee2015,BartelsKrishnamurthyWeniger2015,Linden2015} given the number of pulsars that have already been resolved by the Fermi Large Area Telescope (LAT) \cite{FermiPulsar2013,CholisHooperLinden2014}.
Another proposal is a burst, a continuous injection, or series of bursts of cosmic-ray injections in the GC \cite{CarlsonProfumo,Petrovic2014}. However, there is debate about whether a working version of this is fine-tuned {\cite{Cholis2015,YangAharonian2016}}.
More exotically, dark matter (DM) particles with masses of about $10-100\ \mathrm{GeV}$ annihilating into a variety of  channels  have also been proposed~\cite{Goodenough2009gk,Hooper:2010mq,hooper,AbazajianKaplinghat2012,AbazajianKaplinghat2013,GordonMacias2013,MaciasGordon2014,Abazajian2014,Daylan:2014}. There is debate \cite{Caloreetal:Taleoftails,AbazajianKeeley2015} as to what extent the DM explanation is consistent with Fermi-LAT observations of dwarf satellites of the Milky Way \cite{Geringer-SamethKoushiappasWalker2015,Fermi-LATDwarfSpheroidal2015,Geringer-Sameth2015}.

Secondary electrons and positrons ($e^{\pm}$) would be injected into the interstellar medium either by an unresolved population of $\sim 10^3$ MSPs  or via DM annihilations, if either of these is responsible for the $\gamma$-ray excess seen towards the GC.  The interaction of such particles with the interstellar radiation field (ISRF), galactic magnetic fields and interstellar gas, would modify not only the energy spectrum but also the spatial morphology of the extended $\gamma$-ray source.

The model prediction from DM annihilation secondaries is discussed in Refs.~\cite{CirelliSerpicoZaharijas2013,Cirelli_cookbook_secondaries}.
The Fermi-LAT constraints on  secondaries from DM annihilations in the GC were considered in \cite{gomez2013,Lacroix:2014,Abazajian2014,Daylan:2014} but  only the spectral changes were included in the likelihood analysis. The authors of Ref.~\cite{Abazajian2015} non-parametrically accounted for different secondary spatial morphologies. They used a 20 cm component to model secondary emissions from bremsstrahlung and a template based on infrared starlight emission to model secondary emission from inverse Compton (IC) scattering. They found both these templates inclusion to be preferred by the data. We will take a more parametric approach which can in principle allow us to examine a greater range of interstellar medium (ISM) models. In the context of the MSP explanation of the GeV excess, Refs.~\citep{Ioka,PetrovicSerpicoZaharijas2015} have investigated  the importance of secondary emission for multi-wavelength analyses as well as established a reliable MSP luminosity function.

In this article, we examine the importance of also including the different spatial morphology of the secondary emission which results from the diffusion of the secondary electrons. This has been done to some extent in Refs.~\cite{CaloreCholisWeniger2015,Caloreetal:Taleoftails,Kaplinghat2015} but there they exclude $\abs{b}<2^\circ$ and they do not use the full likelihood approach provided by the LAT Science Tools. We also examine different methods of determining whether secondaries make a significant difference to a model fit of the GCE.

\section{Models for the gamma-ray emission}
\label{sec:secondaryemission}

We compute the various components of the $\gamma$-ray emission from DM in the region of interest as follows.
The prompt diffuse $\gamma$-ray intensity for annihilation channel $i$ is simply given by integrating the DM density squared over the line of sight (l.o.s.) coordinate $s$ (see e.g.~Ref.~\cite{Cirelli_cookbook}):
\begin{align}
\left. E_{\gamma}^{2}\dfrac{\mathrm{d}n_{i}}{\mathrm{d}E_{\gamma}\mathrm{d}\Omega} \right | _{\mathrm{prompt}} = \dfrac{E_{\gamma}^{2}}{4\pi} \dfrac{1}{2} \left( \dfrac{\rho_{\odot}}{m_{\mathrm{DM}}}\right) ^{2} \left\langle \sigma v \right\rangle _{i} \dfrac{\mathrm{d}N_{\gamma,i}}{\mathrm{d}E_{\gamma}} \nonumber \\
\times \int_{\mathrm{l.o.s.}} \! \left( \dfrac{\rho(\vec{x})}{\rho_{\odot}}\right) ^{2} \, \mathrm{d}s,
\end{align}
where $\rho(\vec{x})$ is the DM density, $\rho_{\odot}$ the DM density in the solar neighborhood, $m_{\mathrm{DM}}$ the DM mass, $E_{\gamma}$ the $\gamma$-ray energy, $\left\langle \sigma v \right\rangle _{i}$ the annihilation cross-section into channel $i$ and $\mathrm{d}N_{\gamma,i}/\mathrm{d}E_{\gamma}$ the $\gamma$-ray spectrum from this final state, taken from Ref.~\cite{Cirelli_cookbook}. $\Omega$ represents the dependence on solid angle. 

To compute the secondary IC and bremsstrahlung $\gamma$-ray emissions, we first need to compute the electron and positron spectrum, taking energy losses and spatial diffusion into account. In a steady state, this reads (see e.g.~Ref.~\citep{Lacroix:2014})
\begin{equation}
\label{electron_spectrum}
\psi_{\mathrm{e},i} (\vec{x},E) = \dfrac{\kappa _{i}}{b(\vec{x},E)} \int_{E}^{E_{\mathrm{max}}} \! \tilde{I}_{\vec{x}}(\lambda_{\mathrm {D}}(E,E_{\mathrm{S}})) \dfrac{\mathrm{d}N_{\mathrm{e},i}}{\mathrm{d}E_{\mathrm{S}}} \, \mathrm{d}E_{\mathrm{S}},
\end{equation}
where $ \kappa_{i} = (1/2) \left\langle \sigma v \right\rangle _{i} (\rho_{\odot}/m_{\mathrm{DM}})^{2} $, $E_{\mathrm{max}} = m_{\mathrm{DM}}$, and the total energy loss term $b(\vec{x},E)$ is the sum of the synchrotron, IC and bremsstrahlung losses and is given in the Appendix. $\mathrm{d}N_{\mathrm{e},i}/\mathrm{d}E_{\mathrm{S}}$ is the number of electrons produced by the hadronization or decay of final state $i$, and we use the values tabulated in Ref.~\cite{Cirelli_cookbook}. The \textit{halo function} $ \tilde{I}_{\vec{x}}(\lambda_{\mathrm {D}}(E,E_{\mathrm{S}})) $ contains all the information on the way the DM profile is reshaped by spatial diffusion, through the diffusion length $ \lambda_{\mathrm {D}} $. The latter represents the distance traveled by a particle produced at energy $ E_{\mathrm{S}} $ and losing energy during its propagation, down to energy $ E $. It is given by (see e.g.~Ref.~\citep{Lacroix:2014})
\begin{equation}
\lambda_{\mathrm {D}}^{2}(E,E_{\mathrm{S}}) = 4 \int_{E}^{E_{\mathrm{S}}} \! \dfrac{K(E')}{b_{0}(E')} \, \mathrm{d}E',
\end{equation}
where $b_{0}$ is the energy loss term at the center, $b_{0}(E) \equiv b(\vec{0},E)$ and $K$ is the diffusion coefficient, for which we make similar assumptions as in Ref.~\cite{Cirelli_et_al_constraints}:
\begin{equation}
K(E) = K_{0} \left( \dfrac{E}{E_{0}} \right) ^{\delta},
\end{equation}
with $K_{0} = 4.46 \times 10^{28}\ \rm cm^{2}\ s^{-1}$, $E_{0} = 3\ \rm GeV$. We take $\delta = 0.33$, corresponding to Kolmogorov turbulence.

We compute the halo function following Ref.~\cite{Lacroix:2014}. We consider a half height of 4 kpc for the diffusion zone.
Note that we treat inhomogeneous energy losses in a simplified way. More specifically, we keep the spatial dependence in the $1/b$ term in Eq.~\ref{electron_spectrum} and in the emission spectrum $P$, but we compute the effect of the diffusion assuming homogeneous losses, equal to the value of the losses at the center, $b_{0}$. This simplification, which allows us to avoid resorting to a full treatment of the inhomogeneous propagation equation, is justified by the fact that the DM profile is sharply peaked at the center, so the profile is essentially reshaped by diffusion according to the parameters of the ISM very close to the GC. Moreover, the spatial dependence of the losses only enters the diffusion length through a square root, so the spatial variation of the flux is dominated by the $1/b$ factor and the emission spectrum. On top of that, the energy losses vary only mildly over the region of interest.

In summary, we used a refined treatment of the secondary fluxes with respect to Ref.~\cite{Lacroix:2014}---where the whole calculation was done assuming homogeneous losses---but using the same accurate treatment of the steepness of the DM profile in the halo functions. For the case of interest this is a good approximation  to the fully inhomogeneous resolution methods used e.g.~in the \textsc{Galprop} code\footnote{\url{http://galprop.stanford.edu/}}, the \textsc{Dragon} code \cite{Evoli2008}, or in Ref.~\citep{Cirelli_cookbook_secondaries}, but more straightforward in terms of computation techniques. We actually obtain approximately the same morphology as in Ref.~\citep{Cirelli:2014lwa} where the flux was computed with \textsc{Dragon} as shown in  Appendix \ref{DRAGONCompare}.

The spectrum $\psi_{\mathrm{e},i}$ is then convolved with the emission spectrum $P = P_{\mathrm{IC}} + P_{\mathrm{brems}}$ (given in the Appendix) to obtain the photon emissivity:
\begin{equation}
j_{i}(\vec{x},E_{\gamma}) = N_{\mathrm{e}} \int_{E_{\mathrm{e}}^{\mathrm{min}}}^{E_{\mathrm{max}}} \! P(\vec{x},E_{\gamma},E_{\mathrm{e}}) \psi_{\mathrm{e},i}(\vec{x},E_{\mathrm{e}}) \, \mathrm{d}E_{\mathrm{e}},
\end{equation}
where $N_{\mathrm{e}} = 2$ accounts for the sum of the electron and positron contributions, since a positron is always simultaneously produced with an electron, and the minimum electron energy is $E_{\mathrm{e}}^{\mathrm{min}} \sim E_{\gamma}$. In practice, bremsstrahlung is subdominant compared to IC. This was not the case in Ref.~\cite{Lacroix:2014}, where we had used a simplified model for the gas density, corresponding to higher bremsstrahlung losses than what is obtained here using the \textsc{GALPROP} maps (see Appendix).

The intensity from secondary emissions is then given by integrating the emissivity over the l.o.s.:
\begin{equation}
\label{secondary_flux}
\left. E_{\gamma}^{2}\dfrac{\mathrm{d}n_{i}}{\mathrm{d}E_{\gamma}\mathrm{d}\Omega} \right | _{\mathrm{sec}}= \dfrac{E_{\gamma}}{4\pi} \int_{\mathrm{l.o.s.}} \! j_{i}(\vec{x},E_{\gamma}) \, \mathrm{d}s.
\end{equation}

The derivation of the secondary e$^+$ and e$^-$ fluxes from MSPs  is essentially the same as DM annihilations  with $E_{\mathrm{max}}$ now given by the maximum injection energy. For pulsars, we use a delta function injection spectrum $\mathrm{d}N_{\mathrm{e}}/\mathrm{d}E_{\mathrm{S}} = \delta (E_{\mathrm{S}} - E_{\mathrm{max}})$ with $E_{\mathrm{max}} = 20\ \rm GeV$, as suggested in Ref.~\cite{Ioka} and discussed in the following. 
We parameterize the amplitude of the secondaries as the ratio ($r$) of the energy of gamma-rays observed from the secondary emission to the energy of gamma rays from the primary emission.
Finally, the prompt $\gamma$-ray emission from MSPs is modelled as a power law with exponential cut-off:
\begin{equation}
\frac{dN}{dE}=K\left({E\over E_0}\right)^{-\Gamma}\exp\left(-\frac{E}{E_{\rm cut}}\right),
\label{eq:expcut}
\end{equation}    
where photon index $\Gamma$, a cut-off energy $E_{\rm cut}$ and a normalization factor $K$ are free parameters.

\section{Data Analysis}
\label{sec:dataanalysis}

\subsection{Data Selection}
\label{subsec:fermiobsevations}

The Fermi-LAT  is a $\gamma$-ray telescope sensitive to photon energies from 20 MeV to more than 300 GeV~\citep{2009ApJ...697.1071A}. In operation since August 2008, this instrument makes all sky observations every $\sim 3$ hours.  The angular resolution of Fermi-LAT depends on the photon energy, improving as the energy increases~\citep{2009ApJ...697.1071A}.

The analysis presented here was carried out with  45 months of observations from August 4, 2008$-$June 6, 2012\footnote{Pass-7 data has been superseded by Pass-8. However, the Galactic diffuse emission model corresponding to Pass-8 is not recommended for analysis of extended sources. Hence, we use the 193 weeks of Pass-7 data that are still available at \url{http://heasarc.gsfc.nasa.gov/FTP/fermi/data/lat/weekly/p7v6/photon/}.  However, preliminary tests found similar results with Pass-8 data} as with using the LAT Pass-7 data. The \texttt{SOURCE} class events and the Instrument Response Functions (IRFs)  \texttt{P7SOURCE\_V6}  were used.

In this study, we selected events within a squared region of $7^{\circ}\times7^{\circ}$ centred on Sgr A$^{\star}$, with energies greater than 300 MeV, and without making any distinction between \textit{Front} and \textit{Back} events. For energies lower than 300 MeV the angular resolution of the LAT is poor and source confusion could introduce a large bias to the analysis, whereas above 100 GeV it is limited by low photon statistics.

The zenith angles were chosen to be smaller than 100$^{\circ}$ to reduce contamination from  the Earth limb. Time intervals when the rocking angle was more than 52$^{\circ}$ and when the Fermi satellite was within the South Atlantic Anomaly were also excluded.

The sources spectra were computed using a binned likelihood technique~\citep{3FGL} with the \textit{pyLikelihood} analysis tool\footnote{\url{http://fermi.gsfc.nasa.gov/ssc/data/analysis/documentation/}}, and the energy binning was set to 24 logarithmic evenly spaced bins. The LAT Science Tools\footnote{http://fermi.gsfc.nasa.gov/ssc/data/analysis/}  v9r33p0 was used.

\subsection{ Analysis Methods}
\label{subsec:fermifit}

The spectral and spatial features of an extended $\gamma$-ray source are inherently correlated. Modifications to the spatial model would distort the source spectra and vice versa~\citep{SpatiallyExtended}. It is therefore necessary to assess the impact of secondary $\gamma$-ray radiation in the fit to the GC. We use a fitting method that is fully 3D (comprising an energy axis for the third dimension) and that self-consistently considers the distinct morphological characteristics of the GC extended source in energy and space.

\subsubsection{Fitting Procedure}
\label{subsubsec:fittingmethod}

The complex spectrum and spatial extension of the extended central source is represented as three constituents: (i) Prompt emission, (ii) IC and (iii) a bremsstrahlung component. For the first case, we use a 2D spatial map given by the square of a generalized Navarro-Frenk-White (NFW) profile with an inner slope of $\gamma=1.2$.  The prompt energy spectrum depends on the 
 final states and the different sources of the
case of interest. The remaining two secondary components are modelled by spatially extended sources as explained in Sec.~\ref{sec:secondaryemission}. Their corresponding spatial templates account for spatial variations in energy and are, in this sense, 3D \textsc{MapCube} maps. All three spatial model components have been appropriately normalized\footnote{The reader is referred to the \textit{Cicerone} \url{http://fermi.gsfc.nasa.gov/ssc/data/analysis/scitools/extended/extended.html} for details.} to input in the LAT Science Tools software package. 

This work utilizes two different fitting methods; a broad-band fit analysis and a bin-by-bin analysis procedure:

\begin{itemize}

\item \textit{Broad-band fit:} The fit to the entire energy range ($0.3-100$ GeV) is executed using a similar approach to that followed by the Fermi team in the analysis of the Crab pulsar in the construction of the 3FGL~\citep{3FGL} catalog. The global best fit for the three-component central source is reached iteratively, keeping fixed the parameters describing the spectral shape of the three different components in every iteration step. The flux normalization of the sources are adjusted in such a way that the flux ratio (predicted by our simulations) between the three components is always maintained for the DM case once the DM mass is fixed and for a given annihilation channel. For the MSP case we just maintain the  IC to bremsstrahlung ratio and then leave the ratio to the prompt emission as a free parameter. In practice, this is accomplished by constructing a grid of $\log(\mathcal{L})$ values versus the flux normalization, where $\mathcal{L}$ represents the likelihood of observing the data given the model. A certain point of the grid is obtained in one iteration$-$which is automated in a dedicated computer cluster as this is a computationally intensive task.  

\item \textit{Bin-by-bin fit:} The framework for this stage of the analysis is inherited from Refs.~\citep{GordonMacias2013,MaciasGordon2014}. 
The importance of this step stems from the fact that it works as a form of data compression, allowing us to take into account the systematic uncertainties in the Galactic diffuse emission model. It also serves as a way to validate the spectral and spatial model fit---in the sense that this guarantees that not only the sources are optimized, but that the predictions of the models are consistent with the data.
In cases where the secondaries are negligible, as in the case of DM annihilation to $b\bar{b}$, then the energy bins generated from a good fitting model, like a log-parabola  spectrum, can be used.  But for non-negligible secondaries we may need to account for the three-component nature of the extended source under scrutiny.  As in Refs.~\citep{GordonMacias2013, MaciasGordon2014}, we split the data in several energy bins and run a maximum likelihood routine at each energy bin using the LAT Science Tools. The three-component source is treated similarly as it was done in the broad-band analysis, except that here, the source spectra are replaced by simple power laws with the spectral slope given by the tangent to the broad-band spectra at the logarithmic midpoint of the energy bin. Again, the flux ratio between the three components is kept fixed at all times and a grid of $\log(\mathcal{L})$ values versus the flux normalization for each bin is constructed. In this case, we also keep the ratio of the secondaries to the prompt emission constant in the MSP case. This is necessary as the bin-by-bin fit does not explicitly incorporate changes to the best fit spatial morphology found in the broad-band fit. Ref.~\citep{MaciasGordon2014} computed the systematic uncertainties in the Galactic diffuse emission model obtaining that these are space and energy dependent and of order 20\%.  For this study we use similar analysis methods and assume the same estimates for the systematic uncertainties.

\end{itemize}

\subsubsection{Other Sources included in the Fits}
\label{subsubsec:sourcemodels}

In the broad-band fit,  the spectral parameters of every source (other than the GCE) within $5^{\circ}$ of Sgr A$^{\star}$ were freed, while in the bin-by-bin analysis, only their amplitudes were varied. We employed all 2FGL~\citep{2FGL} point-sources present in the region of interest plus the standard diffuse Galactic emission \texttt{gal$_{-}$2yearp7v6$_{-}$v0.fits} and the isotropic extra-galactic background model \texttt{iso$_{-}$p7v6source.txt}. 

Since the Fermi data used in this work comprises almost 4 years of data taking, while the 2FGL catalog~\citep{2FGL} was constructed with 2 years of data, we are required to make a search for new point-sources in the region of interest. We used the results of Ref.~\citep{MacGorCroProf2015} where two new faint point-sources were found. 

We also included the GC ridge like emission template mentioned in Sec.~\ref{sec:introduction}.
The 2FGL point sources: ``the Arc'' (2FGL J1746.6-2851c) and ``Sgr B'' (2FGL J1747.3-2825c) are spatially coincident with our GC ridge map template.
 It is possible that these two point sources are the result of the interaction of cosmic rays with molecular gas clouds and are thus an integral part of the Galactic ridge. The template for the Galactic ridge source is obtained from a 20-cm map~\citep{Yusef-Zadeh2013,MaciasGordon2014} and for the spectra we used a broken power law~\citep{MacGorCroProf2015}. In the current article we are interested in the spherically symmetric GCE and so we want the best model for the ridge-like excess emission. Therefore, in this article we do include the Arc and Sgr B point sources as well as the 20cm template. This was also done in \citep{MaciasGordon2014} when the goal was to construct a bin-by-bin analysis for the spherically symmetric GCE.

\section{Models and procedure}
\label{sec:results}

We study a set of well-motivated models for the GeV excess for which the propagation of secondary leptons can contribute appreciably to the total energy spectrum, and the resulting $\gamma$-ray spatial morphology can deviate from that given by the square of a generalized NFW profile\footnote{Although in practice a power-law profile would give approximately the same results.} with an inner slope of $\gamma=1.2$. The cases under scrutiny are:

\begin{itemize}

\item \textit{Model I}: 10 GeV WIMPs self-annihilating democratically into leptons ($\frac{1}{3}e^{+}e^{-}+\frac{1}{3}\mu^{+}\mu^{-}+\frac{1}{3}\tau^{+}\tau^{-}$). Based on a spectral fit to the GCE data, Ref.~\citep{Lacroix:2014} found this to be good fitting model provided that the energy spectrum from secondaries was taken into consideration. 

\item \textit{Model II}: 10 GeV WIMPs self-annihilating into $0.25\mu^{+}\mu^{-}+0.75\tau^{+}\tau^{-}$. The stringent constraints on the $e^{\pm}$ annihilation channel obtained by Refs.~\citep{Bergstrom:2013jra,Ibarra:2013zia,Bringmann:2014lpa} motivate this model. Ref.~\citep{Lacroix:2014} showed these particular branching ratios to be the most adequate mixture of leptons other than $e^{\pm}$, that fits well the GC excess energy spectrum.

\item \textit{Model III}: An unresolved population of order $10^3$ MSPs. These objects can release a significant amount of their total spin-down energy in $e^{\pm}$ winds~\citep{Ioka,PetrovicSerpicoZaharijas2015}. The diffusion of such leptons in the GC environment could not only modify the spatial morphology of the central source at $\sim {\rm GeV}$ energies but also potentially provide distinctive signatures at very high energies ($\sim {\rm TeV}$). Here we focus on the situation where electrons are injected monochromatically (typically at $\sim 20\ \rm GeV$) and are not further accelerated, i.e.~in the absence of a shock region, as discussed in Ref.~\cite{Ioka}.

\end{itemize}

\begin{table*}[ht!]
\begin{ruledtabular}
\begin{tabular}{r|rrr|rrr|rrr}
\centering
Model & \multicolumn{3}{c|}{spectrum, prompt only}  & \multicolumn{3}{c|}{spectrum, prompt$+$secondaries} & \multicolumn{3}{c}{ spectrum+spatial, prompt$+$secondaries}   \\ \hline  
           &$\chi^2$&dof&p-value&$\chi^2$&dof&p-value&$\chi^2$&dof&p-value\\ \hline
 I &37.9 & 11& $8\times10^{-5}$& 30.4&11&$1\times10^{-3}$&16.6&10&$8\times 10^{-2}$\\
  II &34.0 &11 & $4\times 10^{-4}$&26.4 & 11&$ 6\times 10^{-3}$&29.4&10&$1\times 10^{-3}$\\
 III  & 11.9&9 &$2\times10^{-1}$ & 11.0&8&$2\times10^{-1}$&11.0&8&$2\times 10^{-1}$
\end{tabular}
\end{ruledtabular}
\caption{\label{tab:chisquares} Results of the spectral (bin-by-bin) analyses performed on the Fermi GeV excess emission as explained in Sec~\ref{subsubsec:fittingmethod}. In the ``spectrum, prompt only'' and ``spectrum, prompt$+$secondaries'' columns  the secondary emission is assumed to have the same morphology as the primary emission and the 
bins to be fitted to were obtained from \cite{MaciasGordon2014}.}
\end{table*}

\begin{table*}[ht!]
\begin{ruledtabular}
\begin{tabular}{l|r|r|r}
\centering
Model & 
{\rm TS}$_{\rm base+prompt+sec}-$TS$_{\rm base}$& dof$_{\rm base}-$dof&TS$_{\rm base+prompt+sec}-$TS$_{\rm base+prompt}$\\  \hline 
Base 
&  0 & 0 & --\\ \hline
 I   (10 GeV, $\frac{1}{3}e^{+}e^{-}+\frac{1}{3}\mu^{+}\mu^{-}+\frac{1}{3}\tau^{+}\tau^{-}$ WIMPs)& 435.1 & 1&101.7 \\
 II  (10 GeV, $0.25\mu^{+}\mu^{-}+0.75\tau^{+}\tau^{-}$ WIMPs)& 343.8 & 1&6.5 \\
III  (MSPs, exponential cut-off) & 512.0  & 4&41.7 \\
\end{tabular}
\end{ruledtabular}
\caption{\label{tab:LogLikelihoods} Results of the broad-band fits to the GCE as explained in Sec~\ref{subsubsec:fittingmethod}. Different models for the GCE in the 300 MeV--100 GeV energy range are listed. Each model includes the base model and the extra prompt and secondary (sec) emission.
The DM models each require one degree of freedom (dof) for the cross-section. The exponential cut-off model requires three parameters for the prompt spectrum and one for the prompt to secondary ratio if included. The spatial morphology of the prompt emission was modelled with a square of a generalized NFW profile with an inner slope of $\gamma=1.2$ \citep{GordonMacias2013}. Spatial templates for the IC and bremsstrahlung components as well as their respective spectra were obtained from our calculations discussed in Sec~\ref{sec:secondaryemission}.}
\end{table*}

\begin{figure*}[p!]
\begin{center}
\begin{tabular}{cc}
\centering
\includegraphics[width=0.5\linewidth]{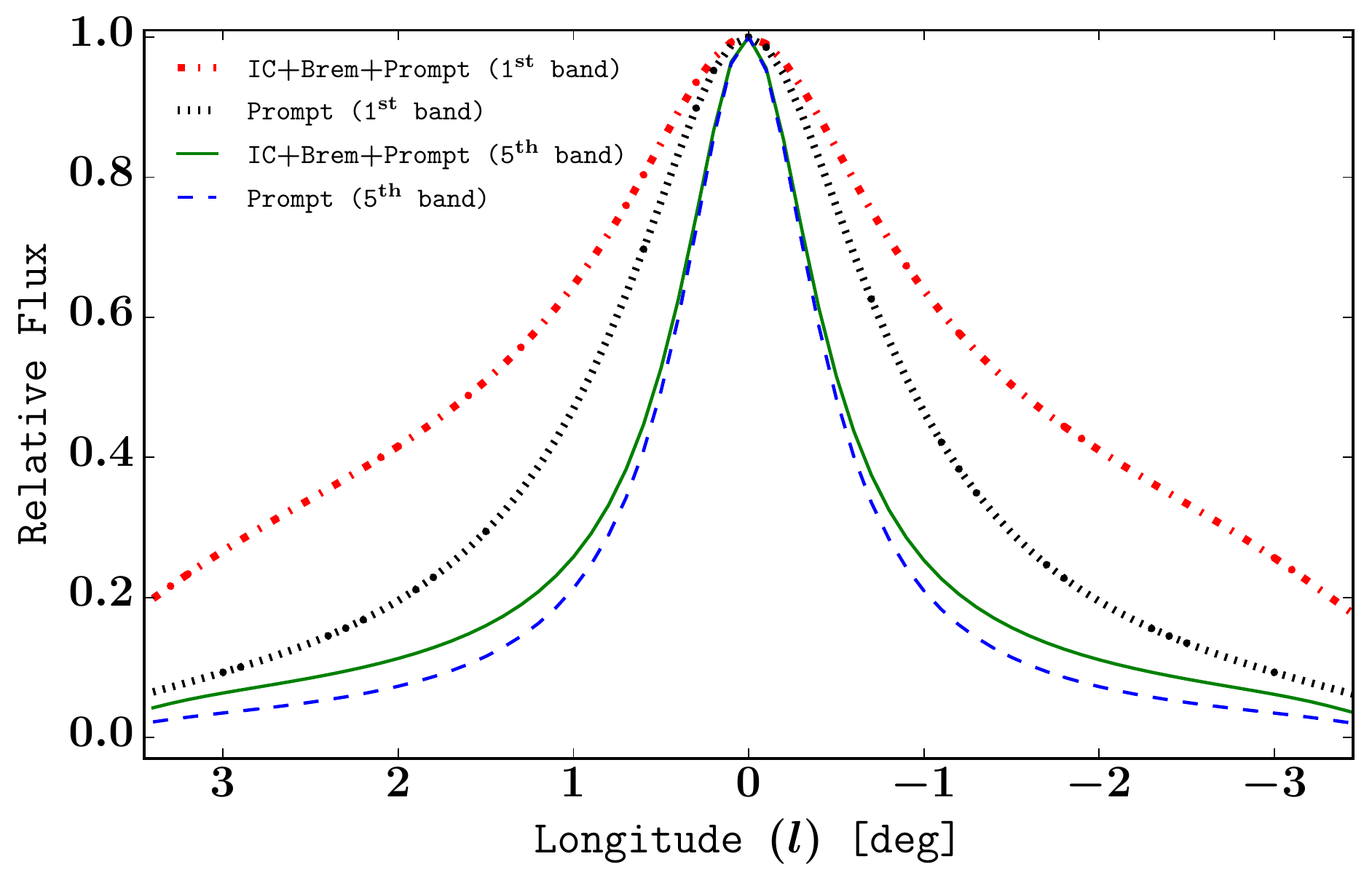} & \includegraphics[width=0.5\linewidth]{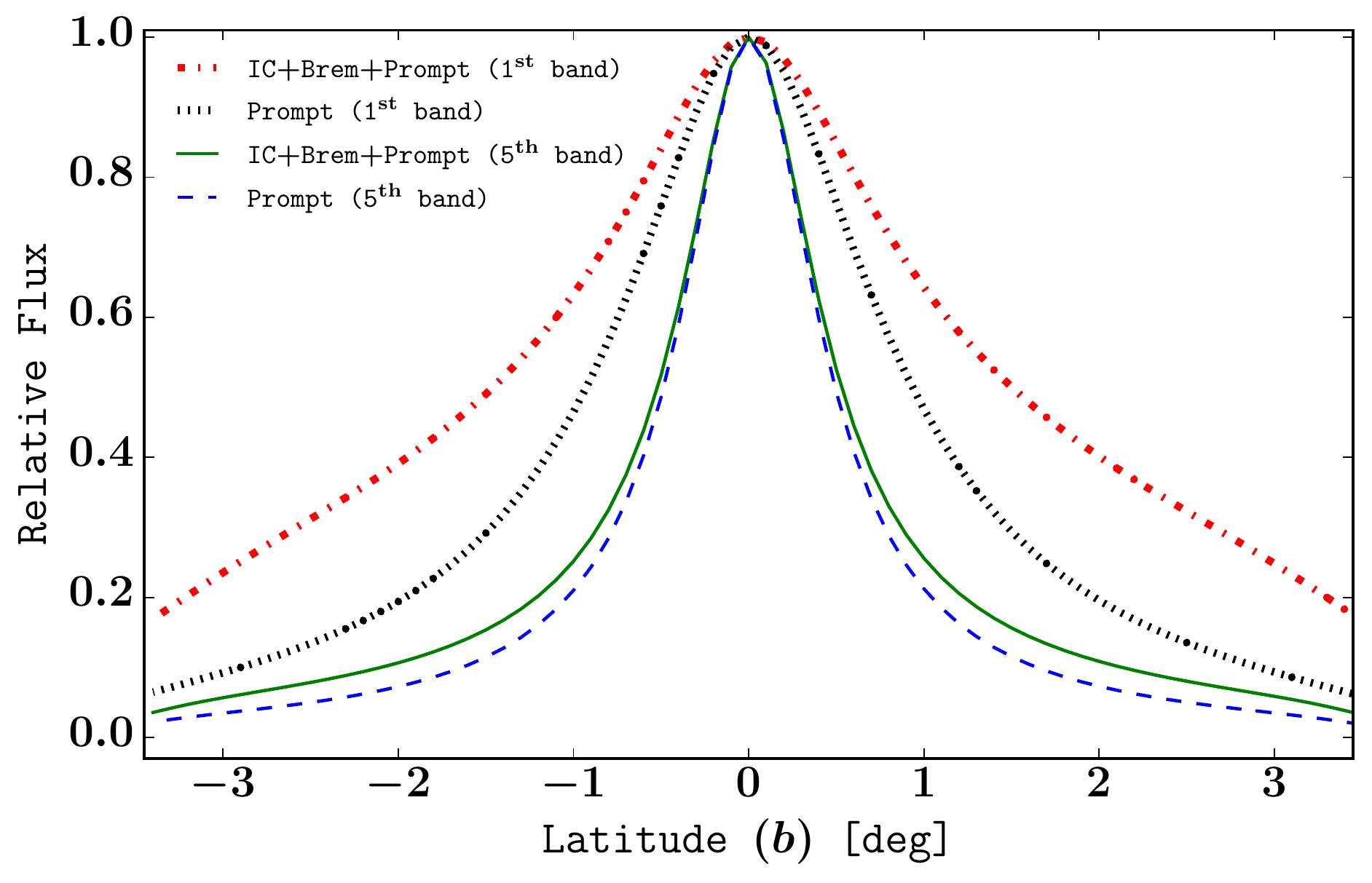}\\
\includegraphics[width=0.5\linewidth]{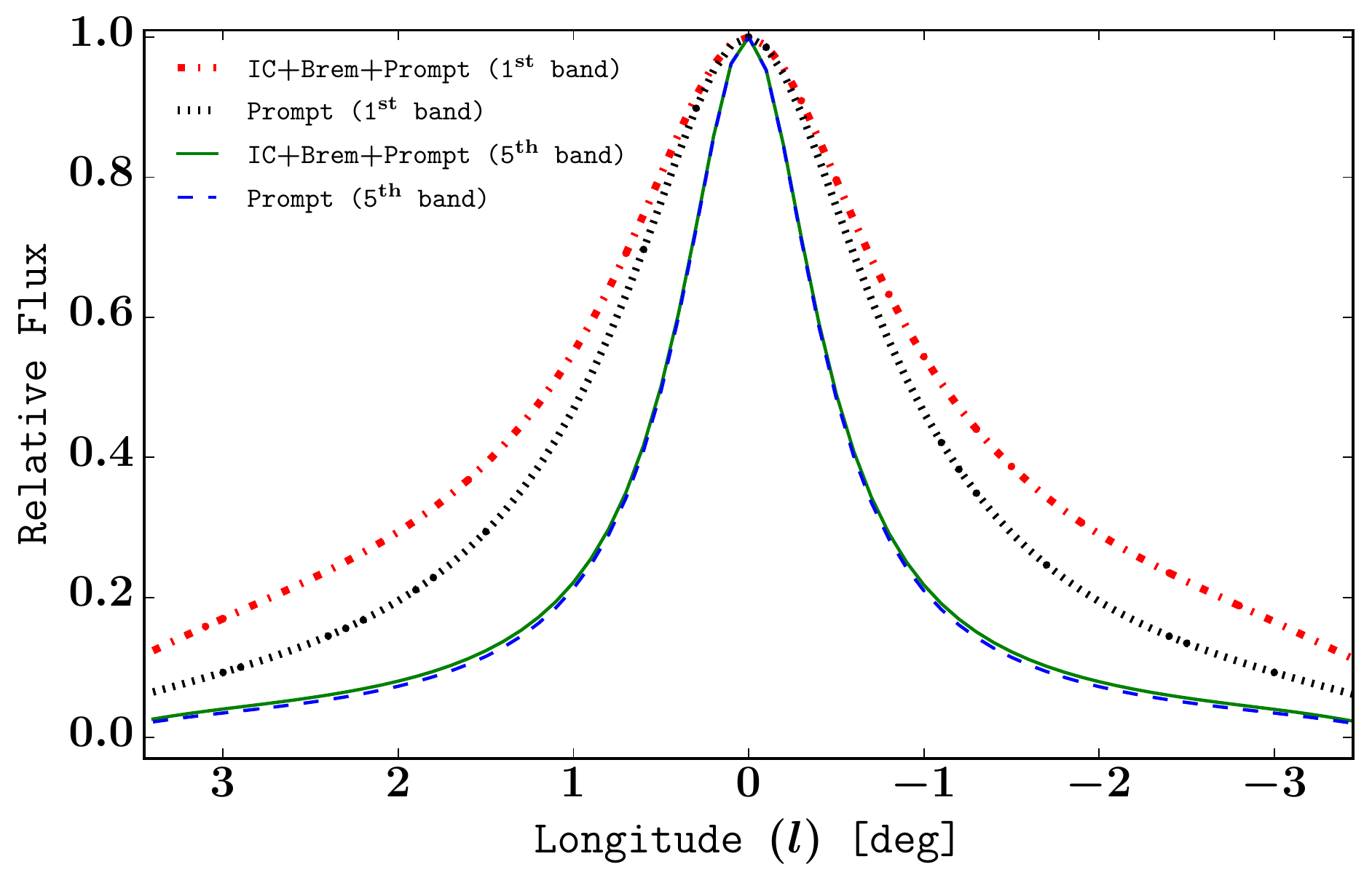} & \includegraphics[width=0.5\linewidth]{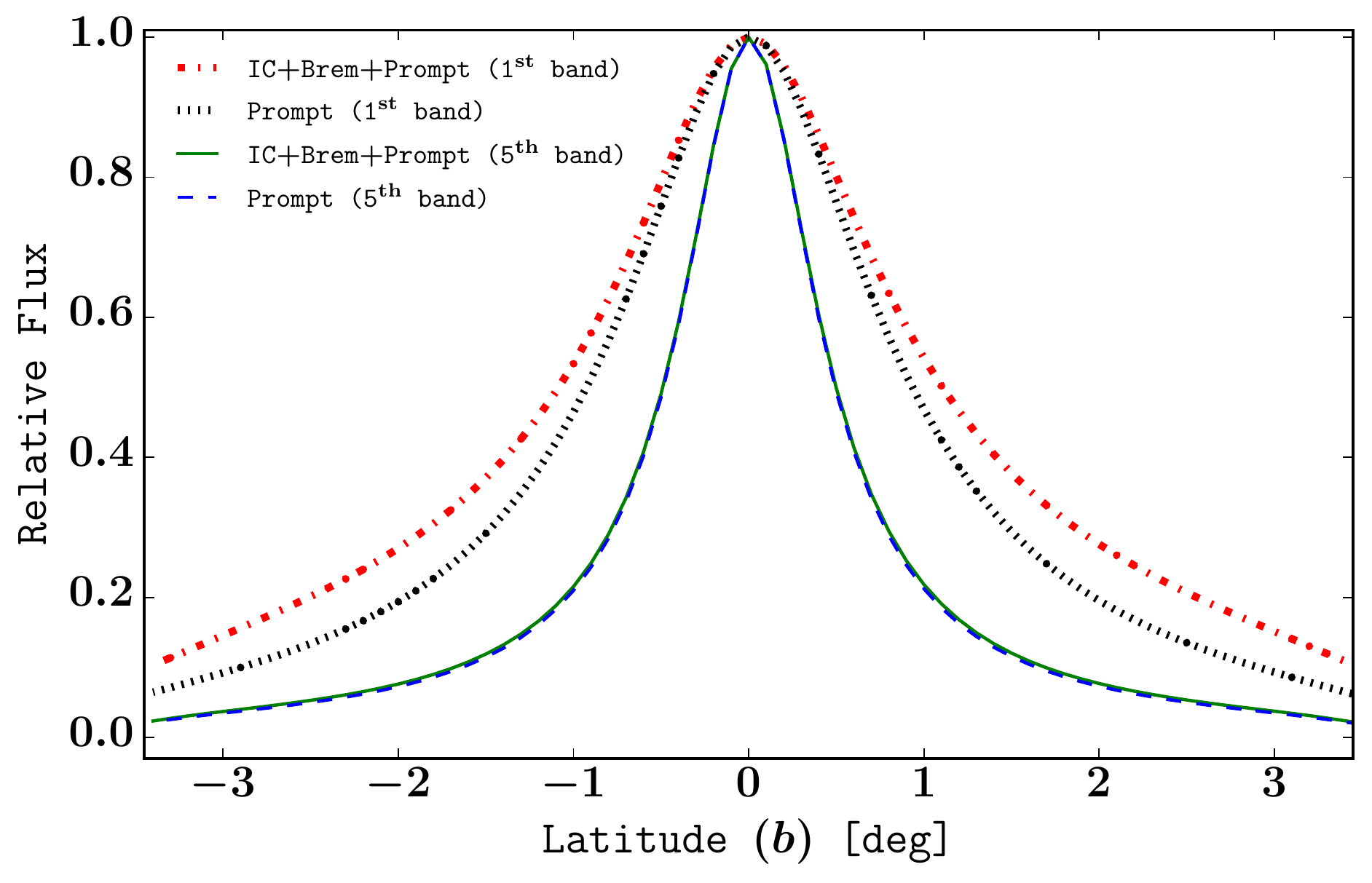}\\
\includegraphics[width=0.5\linewidth]{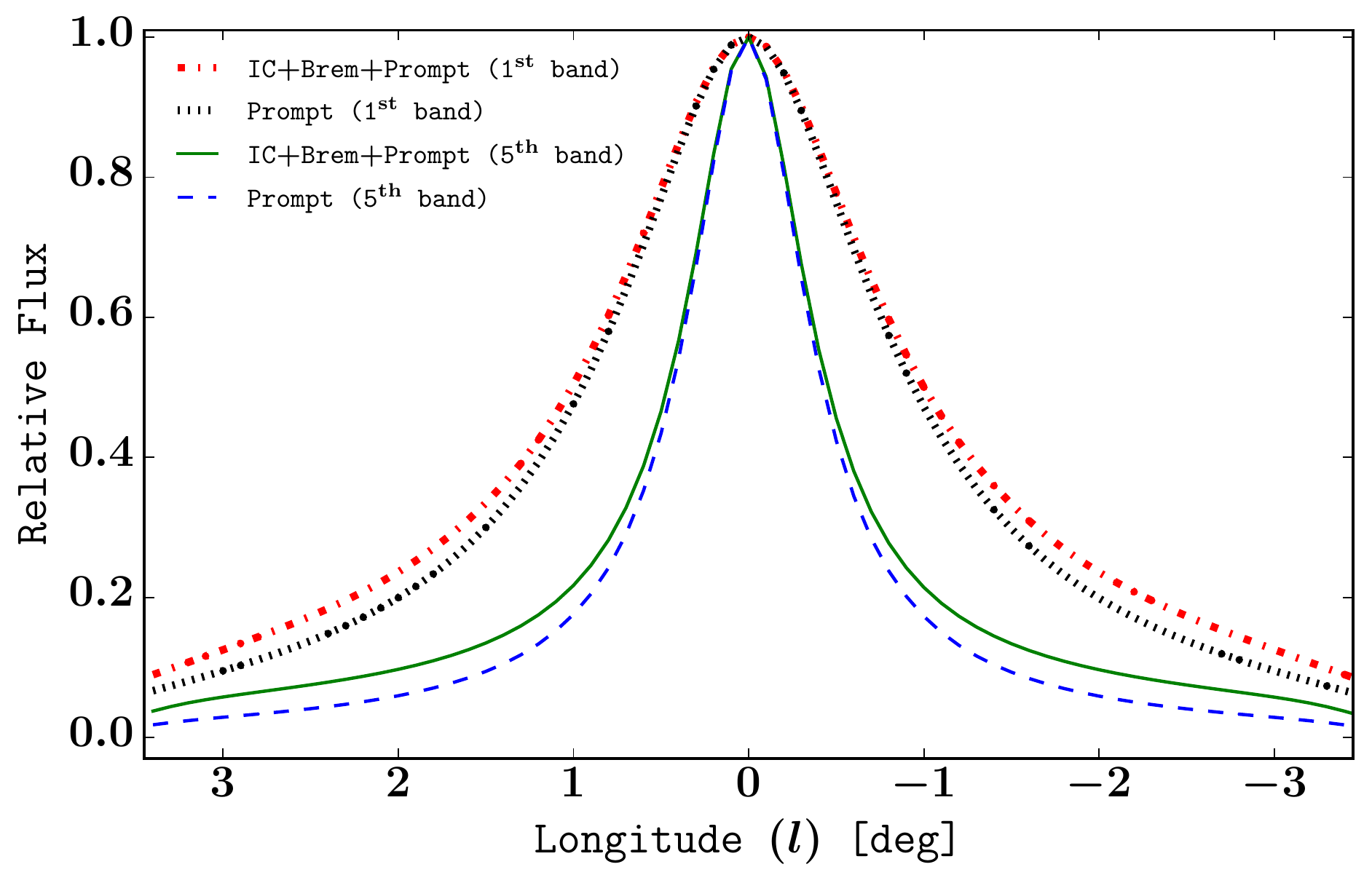} & \includegraphics[width=0.5\linewidth]{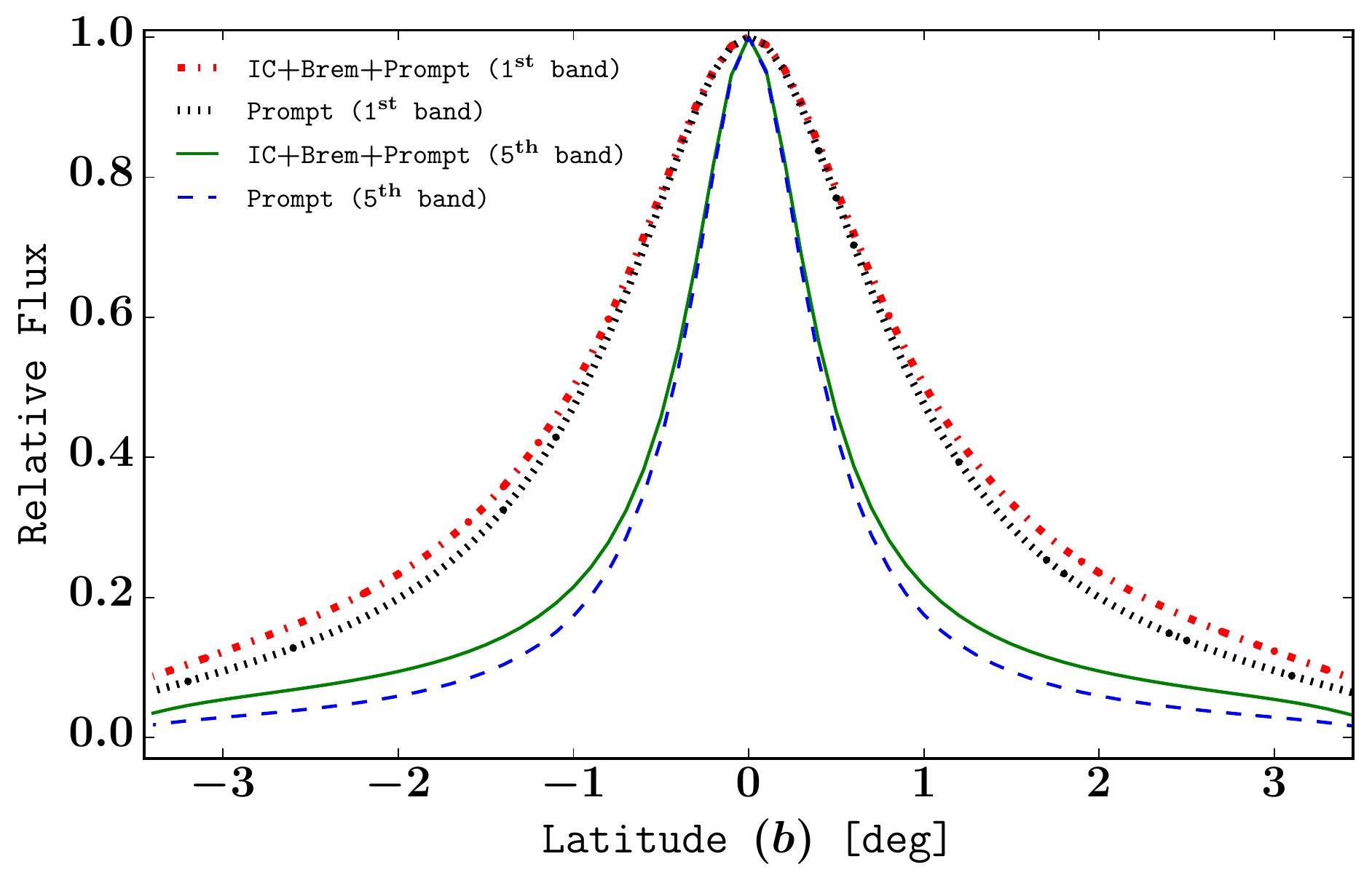}      
\end{tabular}
\caption{ \label{fig:profile}  Spatial brightness profiles of the best fit GeV excess source associated to \textit{Model I} (top), \textit{Model II} (middle), and \textit{Model III} (bottom). Fermi-LAT energy-dependent beam smoothing is included. The profiles are shown in two different energy bins; the $1^{\rm st}$ bin refers to the energy range $0.30$--$0.40$ GeV (thick dotted lines) while the $5^{th}$ one to $0.97$--$1.29$ GeV (thin lines). At each energy bin we present the total emission (IC$+$Bremsstrahlung$+$Prompt) and the prompt emission for comparison. Profiles correspond to $b=0^{\circ}$ (left) and $l=0^{\circ}$ (right). Fluxes are normalized to the maximum for display purposes. See Fig.~\ref{fig:spectrum} for details on the corresponding bin-by-bin analysis.
}
\end{center}
\end{figure*}

\begin{figure*}
\begin{center}
\begin{tabular}{cc}
\centering
\includegraphics[width=0.5\linewidth]{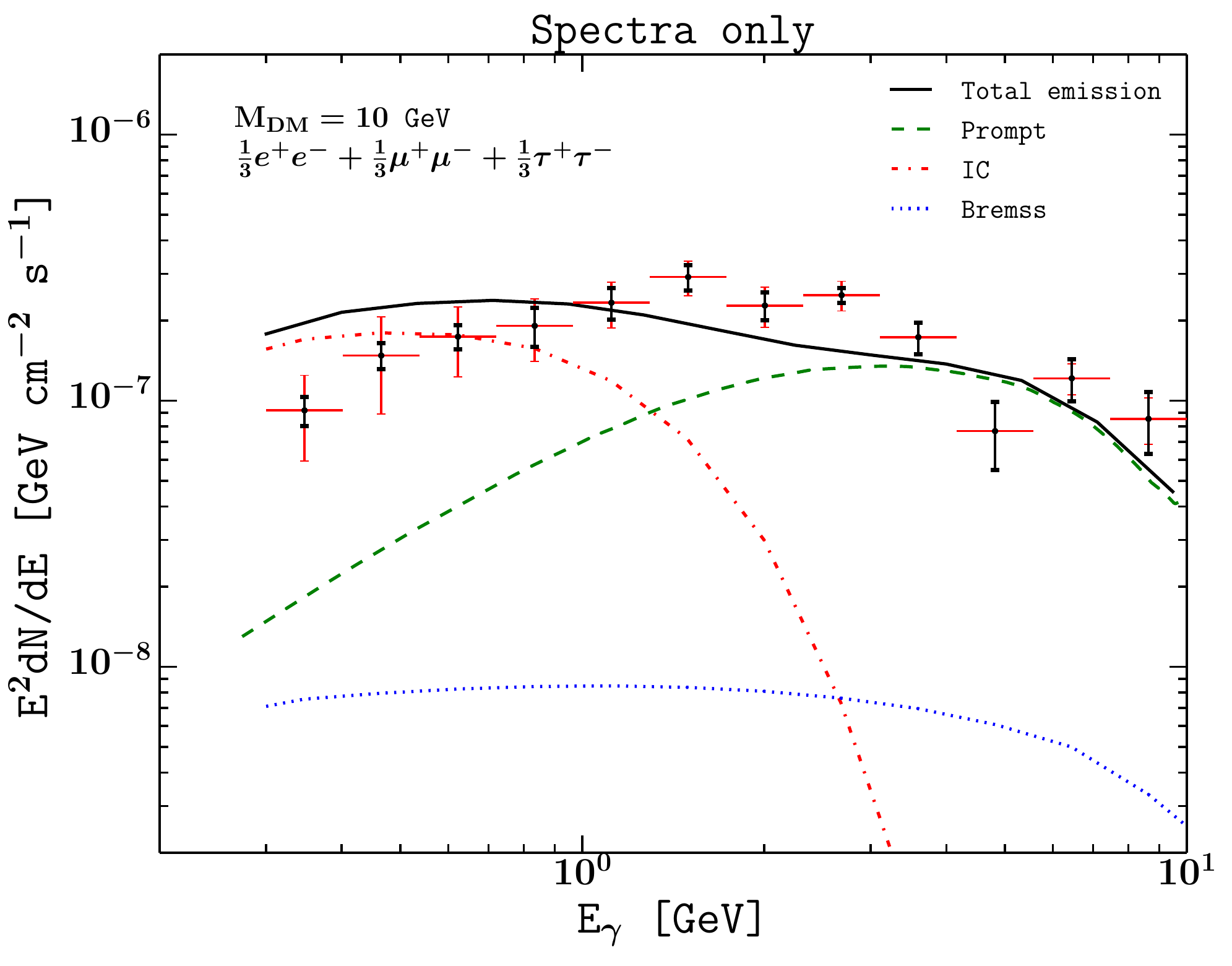} & \includegraphics[width=0.5\linewidth]{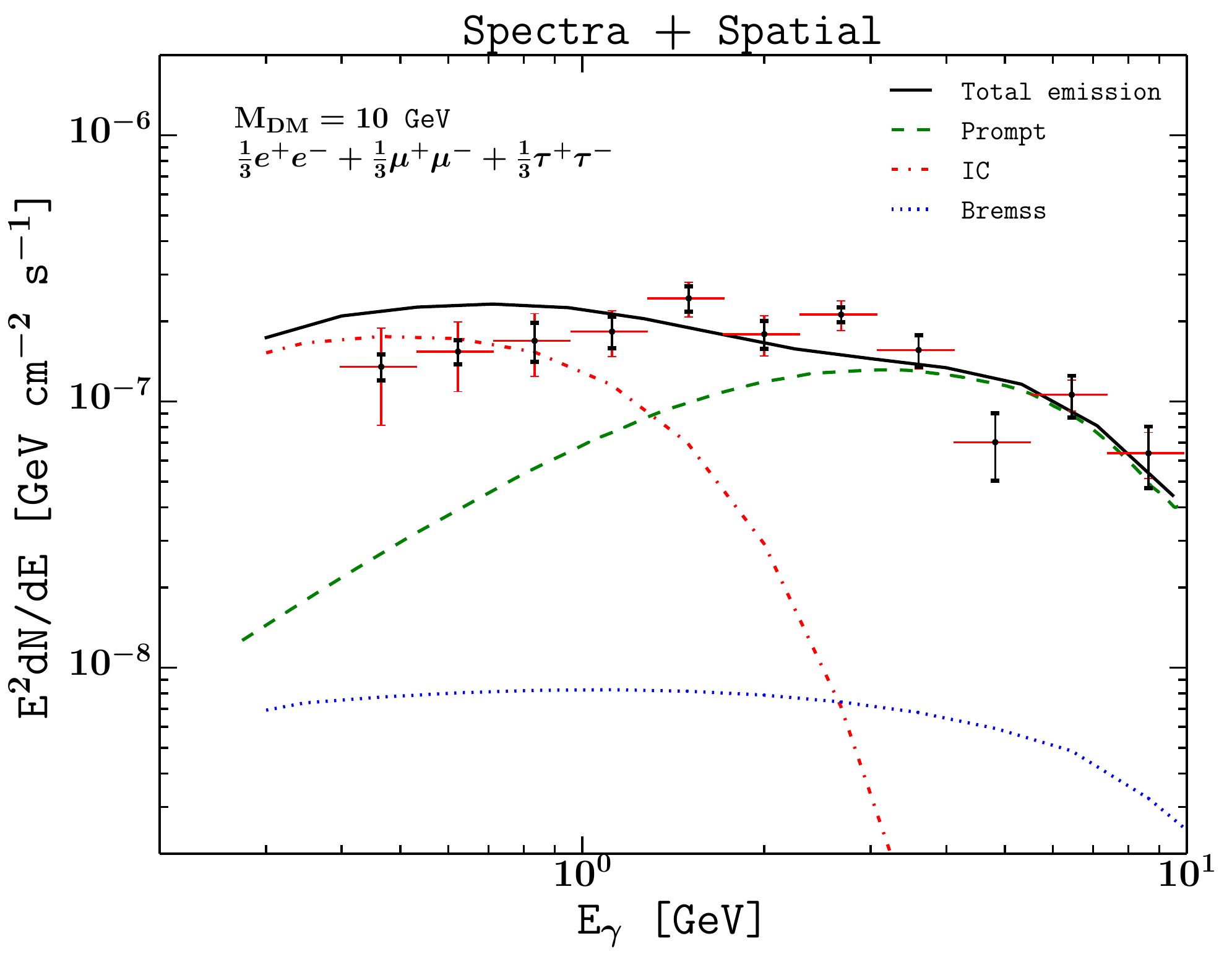}  \\
\includegraphics[width=0.5\linewidth]{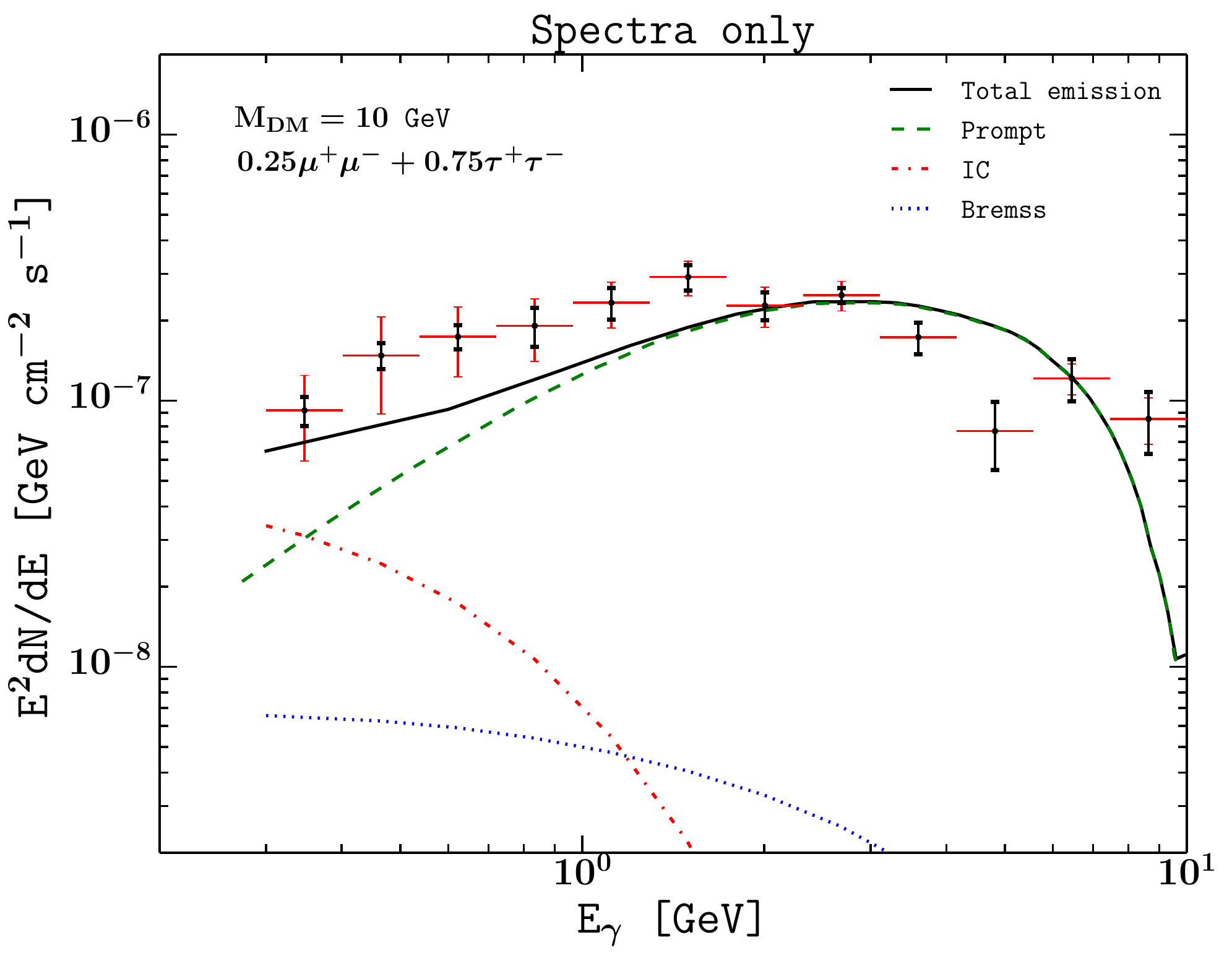} & \includegraphics[width=0.5\linewidth]{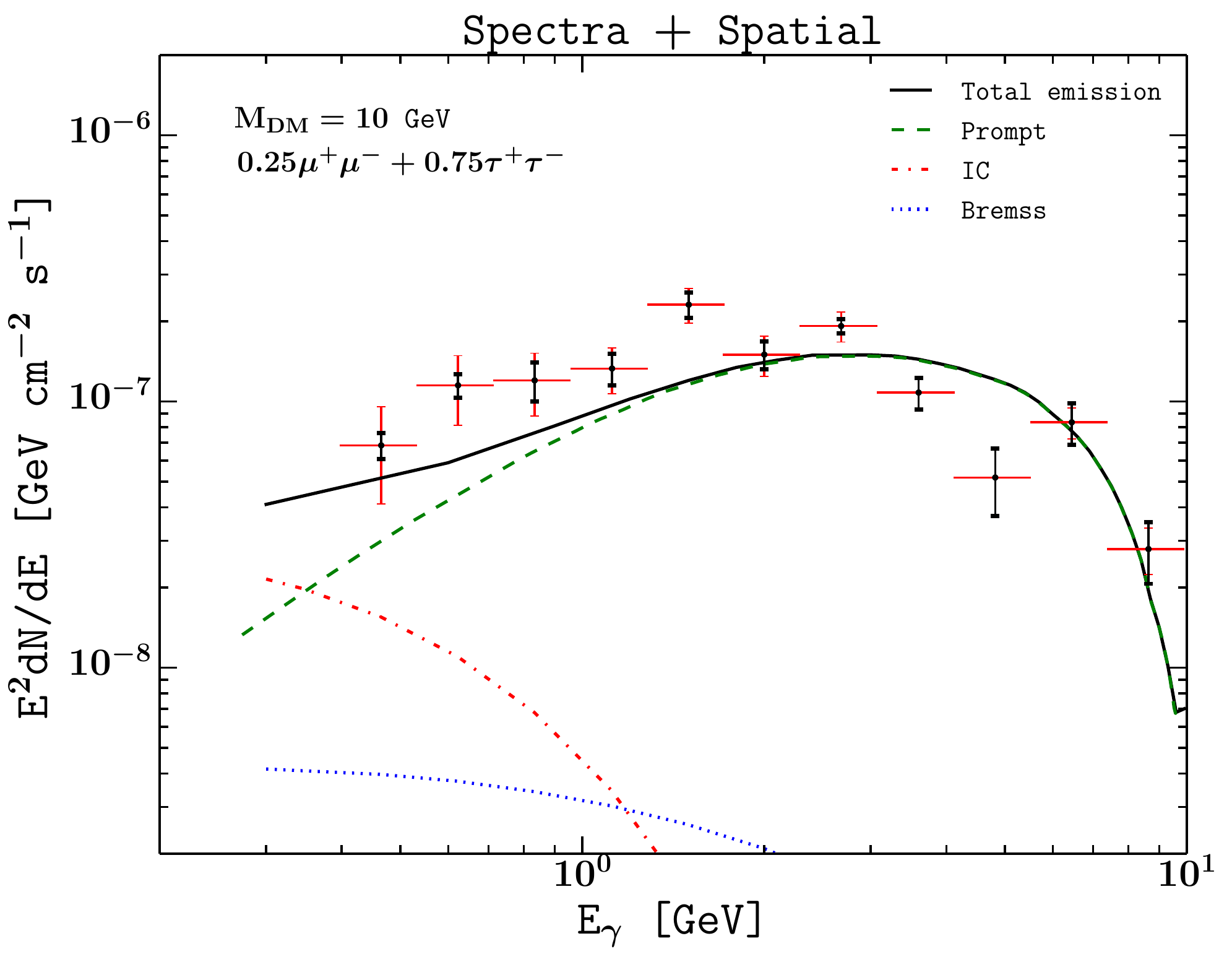}  \\
\includegraphics[width=0.5\linewidth]{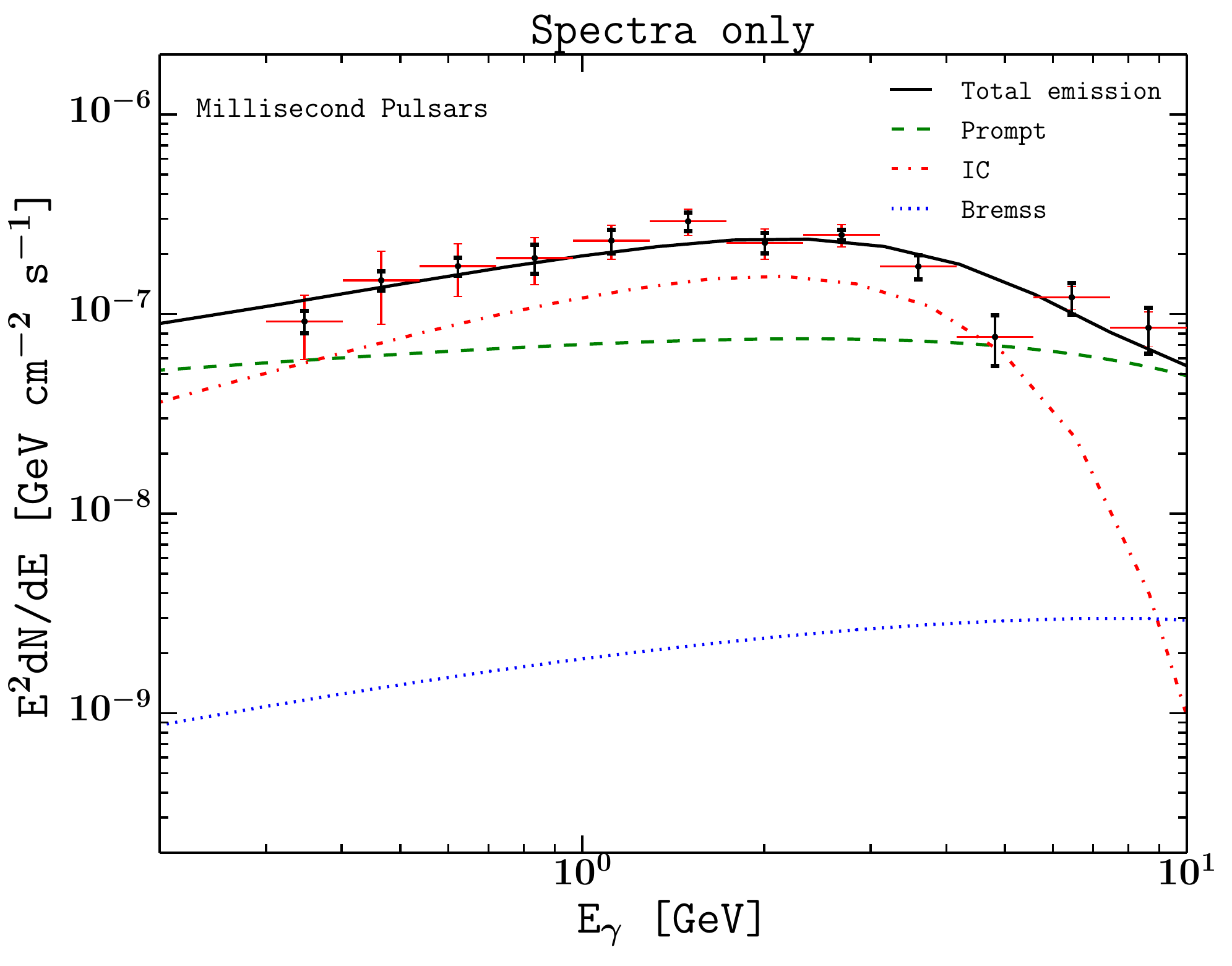} & 
\includegraphics[width=0.5\linewidth]{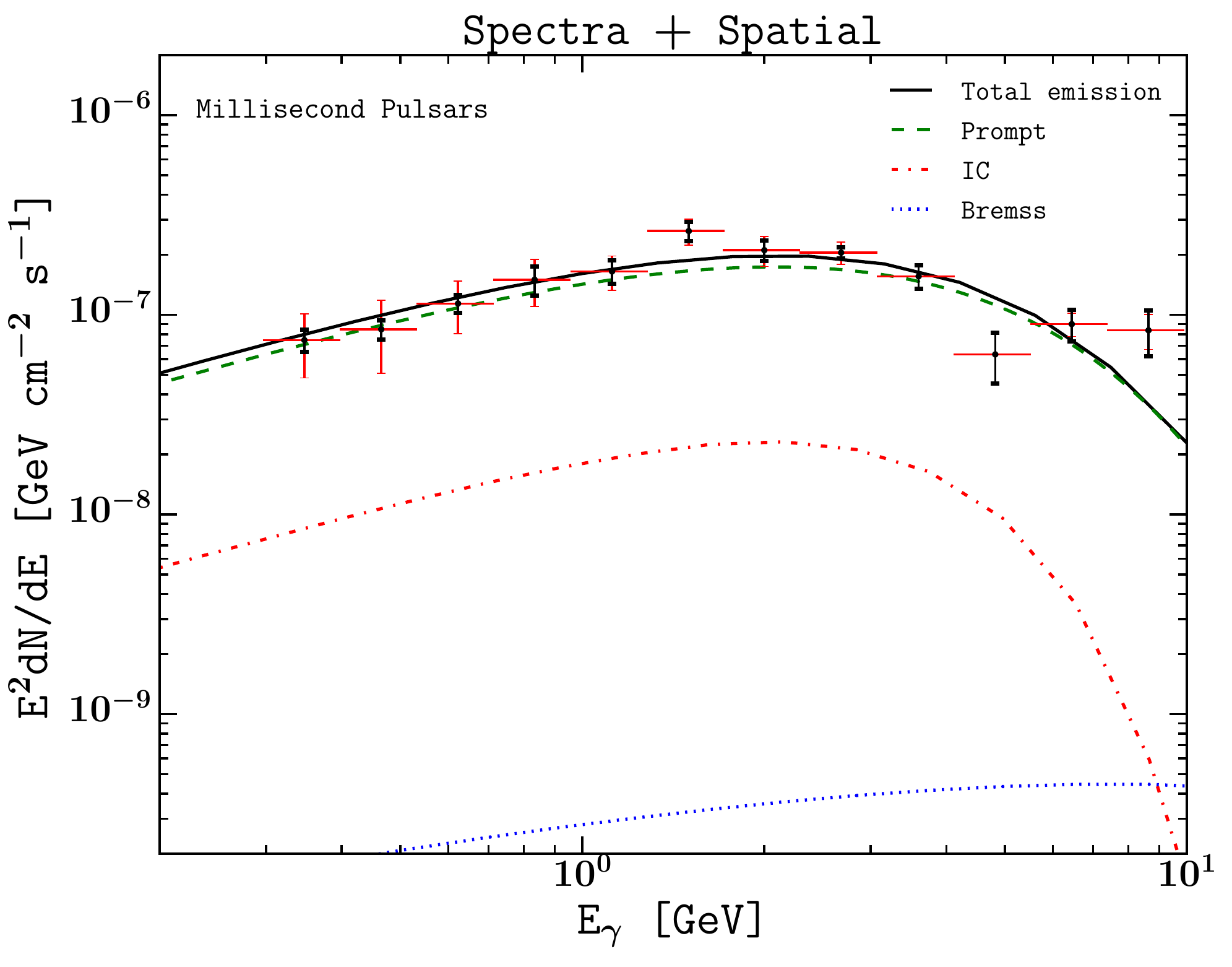}  
\end{tabular}
\caption{ \label{fig:spectrum}  Spectral energy distribution for the GCE modelled with Spatial brightness profiles of  \textit{Model I} (top), \textit{Model II} (middle), and \textit{Model III} (bottom).  IC and Bremss stand for inverse Compton and bremsstrahlung emission respectively. Black and red error bars refer to the LAT ($1\sigma$) statistical and systematic errors, respectively. The fit and plot only consider energy bins with ${\rm TS}\ge 1$. \textsf{Left Panel:} shows the results of a bin-by-bin analysis when the secondaries' different morphologies were not accounted for in determining the bins. \textsf{Right Panel:} Displays the results of the bin-by-bin analysis when the full spectral and spatial information from secondaries was considered.}
\end{center}
\end{figure*}

We start off by using a pure spectral analysis for comparison with previous results from the literature, e.g.~those obtained in Ref.~\citep{Lacroix:2014}. Then, to determine whether a new model component is required, and more importantly to assess the actual importance of secondaries, we perform a 3D broad-band fit to evaluate the value of the test statistic ${{\rm TS}}=2\ln ({\cal L}_{\rm new}/{\cal L}_{\rm old})$, where ${\cal L}$ is the maximum likelihood and the subscript indicates whether or not the new parameters are included. In the case of DM the new parameter corresponds to $\left \langle \sigma v \right\rangle$. For the DM models we consider that the ratio of primary to secondary emission is fixed by the underlying theory's annihilation channels and the assumed ISM parameters. In the MSP case we allow the ratio of primary to secondary emission to be a free parameter and we use the exponential cut-off for the MSP primary spectrum. Crucially the spatial and spectral aspects of the prompt and secondary emission are accounted for. The distinct morphologies of the secondary emissions are illustrated in Fig.~\ref{fig:profile}.

Based on the examination of the sources near Cygnus, Orion and molecular clouds, the Fermi collaboration~\citep{2FGL} stipulated that depending on the intensity of the diffuse background, sources near the galactic ridge need to have ${\rm TS} \gg 25$ to not be considered as simply corrections to the DGB model. A new source would need to have a ${\rm TS}\ge 80$ to be seriously considered for a multi-wavelength search and so we adopt that value as our necessary threshold for a model to explain the GCE. This criterion is based on 4 new parameters and if a source only has one new parameter, an equivalent p-value threshold is obtained by requiring TS$\ge68$.

We can assign a TS for a model's secondary emission by comparing the best fit likelihood with and without the secondaries included. For models which have ${\rm TS}\ge 68$ secondary emission, we proceed to perform a bin-by-bin analysis, to check the consistency of the results as explained in Sec.~\ref{subsubsec:fittingmethod}. For the DM cases, there was only one degree of freedom,  $\left \langle \sigma v \right\rangle$,  in both the broadband and bin-by-bin fit. In the MSPs and log-parabola case the three parameters of  the primary spectrum and the ratio of secondaries to primaries are allowed to vary in the broadband fit. If the secondary and primary morphology is assumed to be the same, then a pure spectral fit, to a previously evaluated primary only bin-by-bin spectrum, can be done with the MSP secondary to primary ratio allowed to vary. However, if the distinct secondary morphology is accounted for in the bin-by-bin fit, only the overall normalization of the total MSP model spectrum is allowed to vary. The other three parameters had to remain fixed, to the broadband best fit values, so as to preserve the underlying spatial morphology which is fixed in the bin-by-bin case.

\section{Results and discussion}
\label{sec:discussion}

Table~\ref{tab:chisquares} summarizes the results of a spectral analysis. The results are plotted in Fig.~\ref{fig:spectrum}. For the LHS panels we used the bins from \cite{MaciasGordon2014} which were generated with a primary-emission only model. To further assess the need for secondaries, once the actual spatial morphology of the secondary emission was taken into account, we performed a 3D broad-band analysis, as described in Sec.~\ref{subsubsec:fittingmethod}. The results for the broadband analysis are shown in Table~\ref{tab:LogLikelihoods}. We also performed  3D bin-by-bin analyses to assess the importance of systematic uncertainties introduced by the spatial morphology. In the RHS panels of Fig.~\ref{fig:spectrum}, the actual secondary emission spatial profiles were used to generate the bins, as explained in Sec.~\ref{subsubsec:fittingmethod}. Note that on the RHS there is one less significant (${\rm TS}\ge 1$) bin compared to the LHS for \textit{Model I} and \textit{Model II}.

\subsection{\textit{Model I}, democratic leptons}

Table~\ref{tab:chisquares} shows that for \textit{Model I}, the fit p-value is improved to the $10^{-3}$ threshold when including secondaries and assuming that their morphology is the same as the morphology of the prompt emission. However, the improvement is significantly above that level once the distinct spatial morphology of the secondaries is accounted for. Table~\ref{tab:LogLikelihoods} shows that the democratic leptons case (\textit{Model I}) has an overall ${\rm TS}\ge68$ and so can be considered as a potential model for the GCE. Moreover, Table~\ref{tab:LogLikelihoods} shows that for this model, secondaries have ${\rm TS}\ge68$, so we conclude that for \textit{Model I}, the need for secondaries is ascertained by the 3D analysis.

Interestingly, using a more non-parametric template-based fitting approach, the authors of Ref.~\cite{Abazajian2015} also found evidence for secondary $\gamma$ rays originating from $\sim 10\ \rm GeV$ electrons, consistent with both the democratic leptons model and the MSP scenario. Their prediction for the democratic leptons bremsstrahlung component was somewhat higher than ours which is likely due to their different approach of extracting the bremsstrahlung contribution from the data and different assumptions about the ISM.

\subsection{\textit{Model II}, no electrons}

As can be seen from Table~\ref{tab:chisquares}, if a spectral-only analysis is performed, the no-electron case (\textit{Model II}) goes from bad-fitting to good-fitting (p-value $\ge10^{-3}$) if secondaries are included. When the distinct morphology of the secondaries is accounted for the goodness of the fit decreases to just above the $10^{-3}$ threshold. From this spectral analysis, we would be led to conclude that secondaries are needed in making \textit{Model II} a good model for the GCE.

When moving to the full 3D broad-band analysis, although we obtain a significant overall TS value, the contribution of secondaries turns out to be negligible, as evidenced by the last column of Table~\ref{tab:LogLikelihoods}. When deciding whether a new model component is needed by the data, evaluating the improvement in the likelihood (via a TS comparison) is a valuable tool. However, there can be cases where the new model component improves some other aspect of the fit which does not significantly change the data likelihood. As we have seen that is what happens in the case of Model II. In that case secondaries do not significantly improve the TS (likelihood) but they do make the  spectral fit acceptable. We therefore argue that our spatial bin-by-bin analysis shows that Model II does require secondaries even if they do not have a significant effect on the broadband TS.

\vspace{-0.45cm}

\subsection{\textit{Model III}, MSPs}

As seen from Table~\ref{tab:chisquares}, the spectral-only analysis would not reveal the need for secondary component for the MSP case (\textit{Model III}) as the p-value is well above the $10^{-3}$ threshold before or after adding the secondaries. 

 As seen from Table~\ref{tab:LogLikelihoods}, the TS of the MSP secondaries is higher than the traditional 25 threshold but not higher than the threshold of 80 needed to be accepted as  a non-correction to the DGB. The best broadband analysis fit value for the secondary to primary gamma-ray ratio was only $r\sim1\%$ and the other parameter values where consistent with the no secondaries case considered in Ref.~\cite{MaciasGordon2014}. As seen from the bottom panels of Fig.~\ref{fig:spectrum}, a larger best-fit value of $r$ was obtained in the spectrum only analysis. But due to large degeneracies with the other parameters, it was less than $3\sigma$ away from the no-secondary case of  $r=0$. Therefore, in this case both the  spectral analysis, and the full 3D broadband analysis show that the data do not require a secondary component. Similar results were found with a log-parabola model.

\section{Conclusions}
\label{sec:conclusions}

In this article, we have illustrated the importance of including the spatial morphology of secondary emission in a self-consistent analysis set-up when evaluating the validity of models for the GeV excess. Our 3D broadband analysis took into account this spatial morphology and by requiring a high TS threshold, we showed that a secondary emission component is required in the democratic lepton  case. This was also confirmed by a spectral analysis which accounted for the different spatial morphologies of the secondaries.

In the no-electron case of \textit{Model II}, the full broadband analysis did not support the need for secondaries. But, a spectral analysis showed that the model fit was below the $10^{-3}$ p-value threshold unless secondaries were included. The TS statistic only tells how much a model is improved by secondaries, but does not take into account how well the overall model fits. This illustrates the need to check model fit in addition to TS improvement. We have shown a spectral approach to evaluating model fit can be adapted to the case where some components of the model have different spatial morphologies.

In future work, we will perform a full likelihood analysis to accurately determine the secondary model parameter uncertainties in the presence of DGB systematics. This will require us to also generate a DGB template which varies with the ISM as at least the IC component should also change when the ISM radiation field is adjusted.

\begin{figure*}
\begin{center}
\begin{tabular}{cc}
\centering
\includegraphics[width=0.5\linewidth]{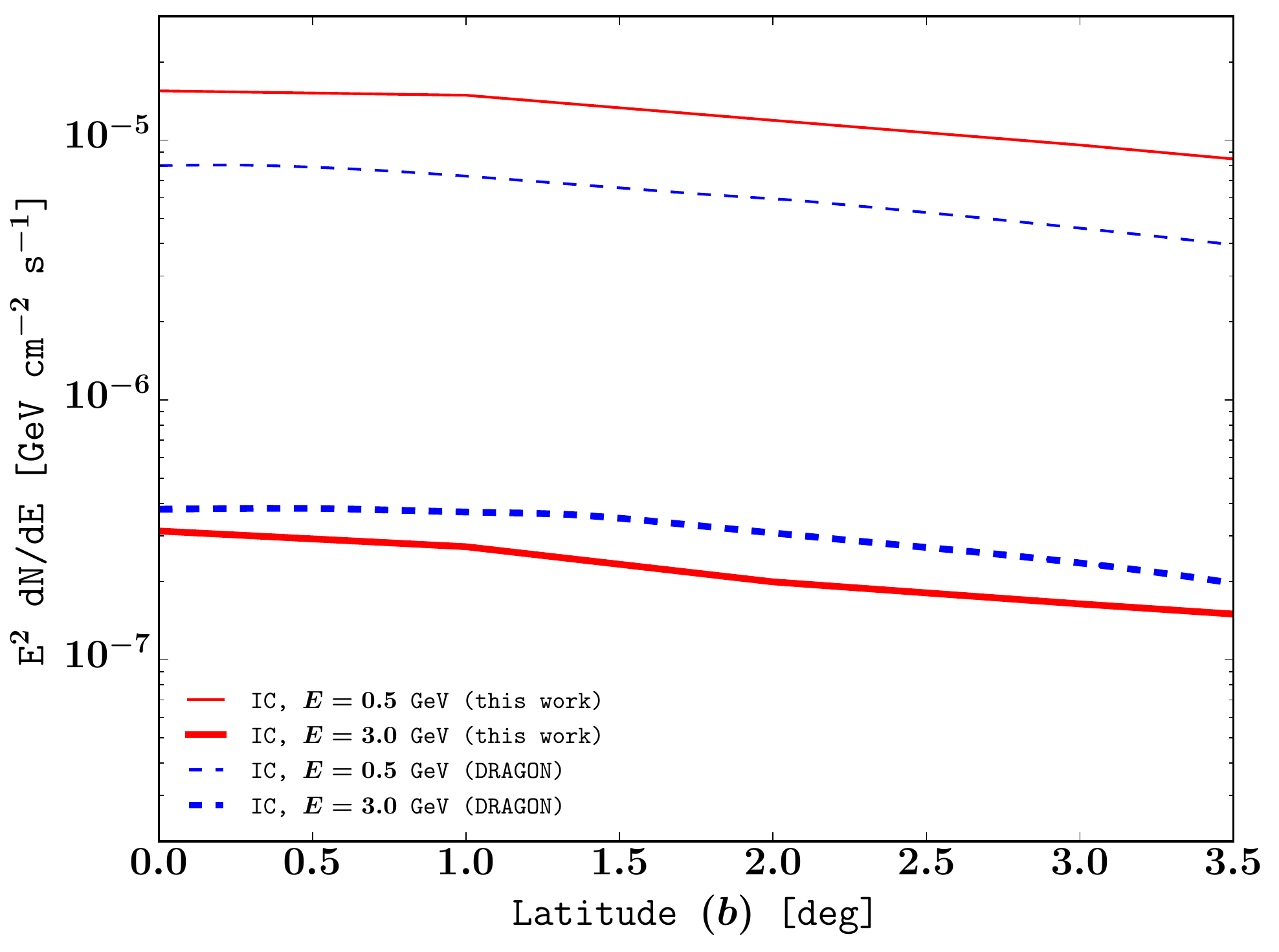} & \includegraphics[width=0.5\linewidth]{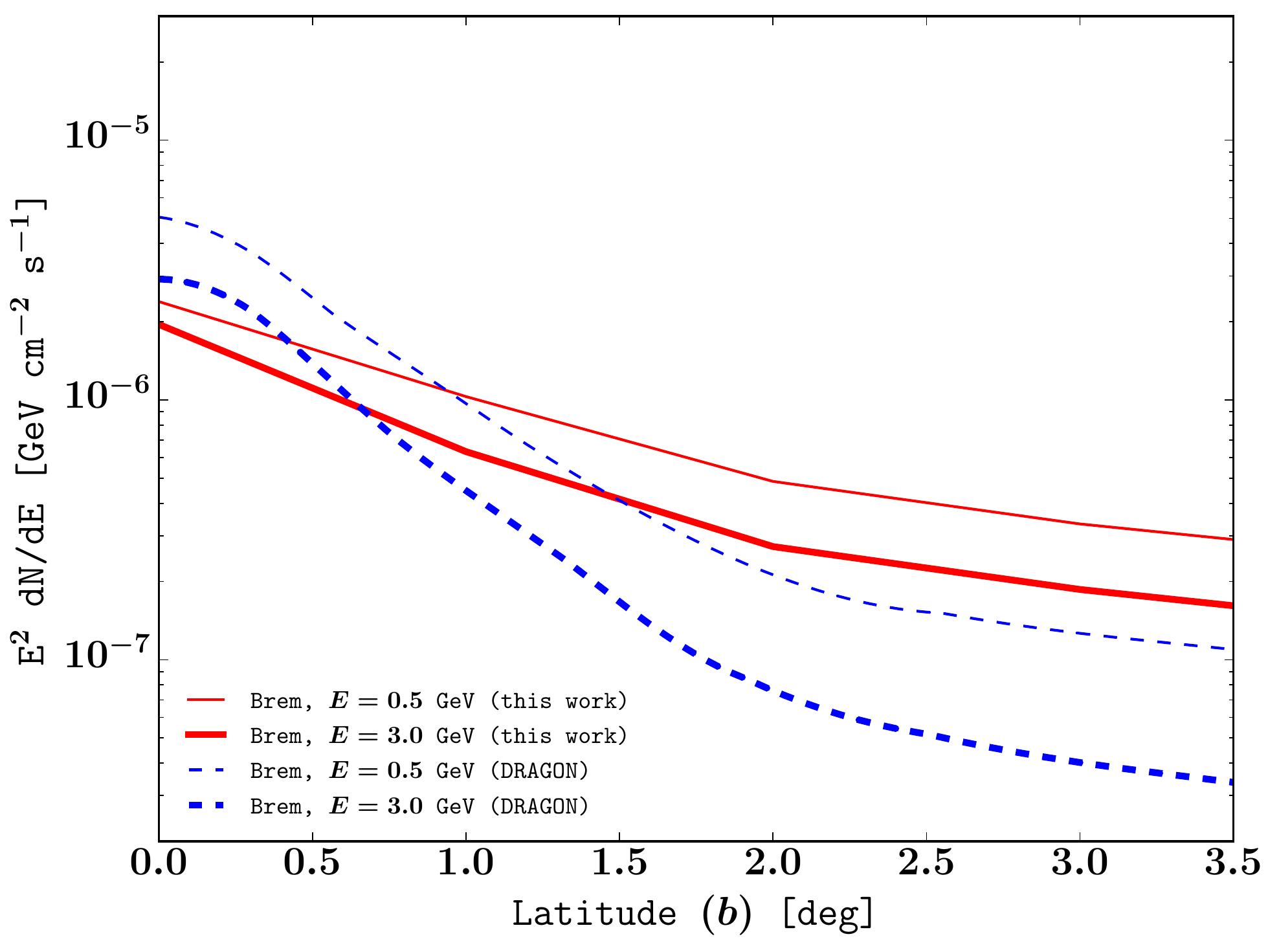}\\
\includegraphics[width=0.5\linewidth]{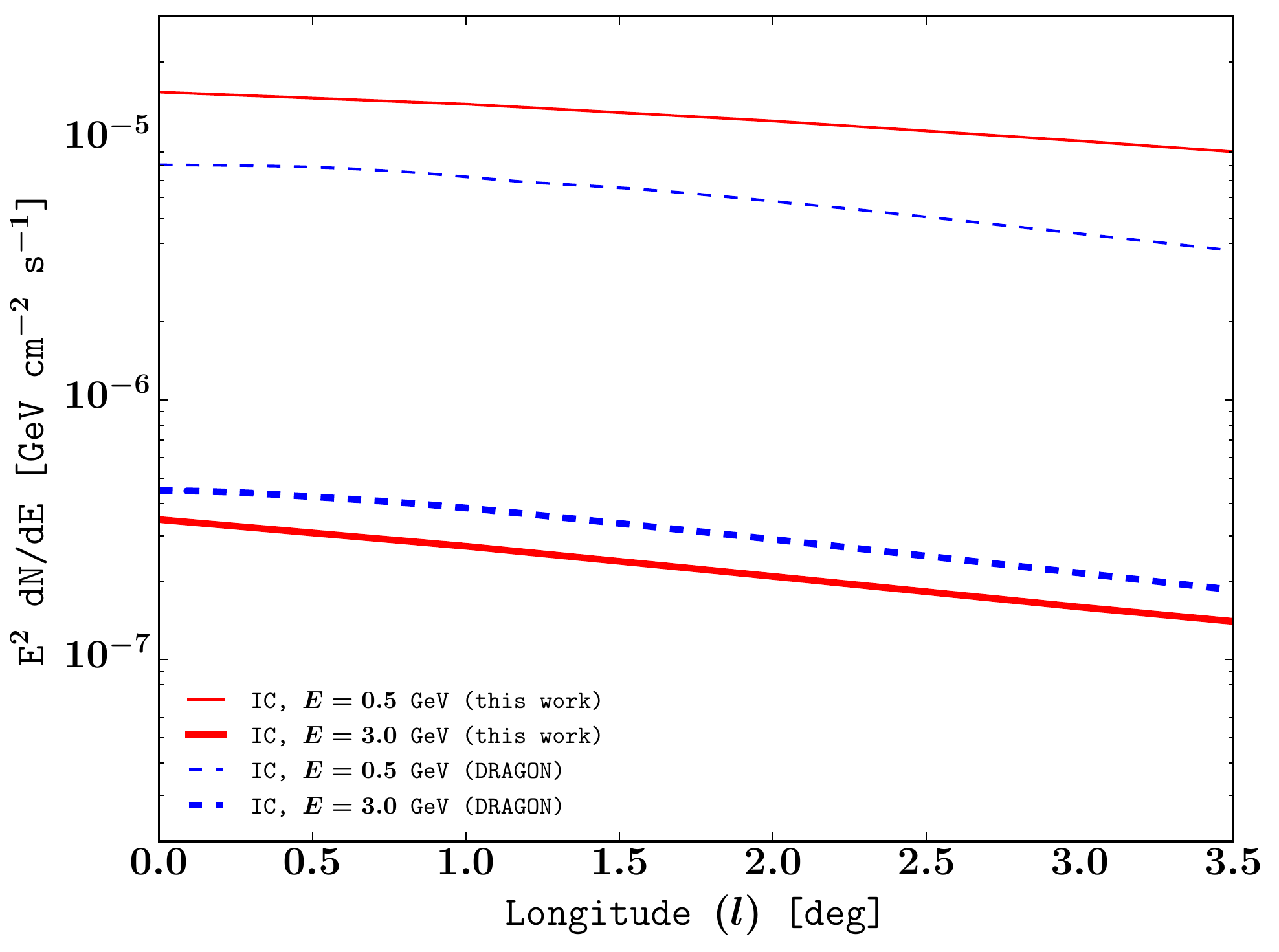} & \includegraphics[width=0.5\linewidth]{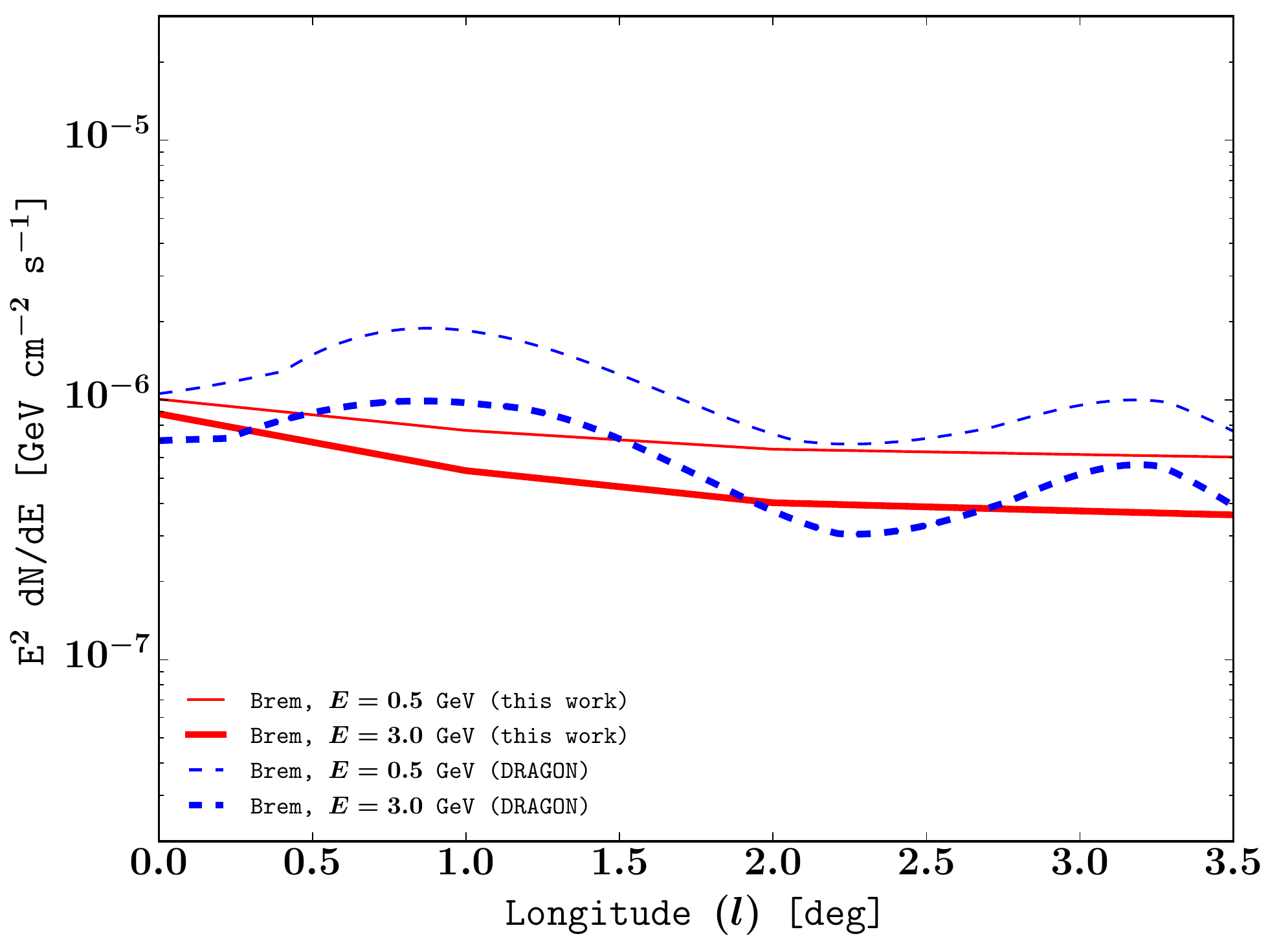} 
\end{tabular}
\caption{ \label{fig:DRAGONcomparison}Comparison of our secondaries with those generated by \textsc{Dragon} for leptons. A model with DM mass of 9.4 GeV, $\left\langle \sigma v \right\rangle=3\times 10^{-26}$~cm$^{3}$s$^{-1}$, and an annihilation channel mix of $20\% e^+e^- + 20\% \mu^+\mu^- + 60\% \tau^+\tau^- $ is used. The \textsc{Dragon} predictions are taken from Fig.~2 of Ref.~\cite{Cirelli_et_al_constraints}.
}
\end{center}
\end{figure*}

\begin{acknowledgements}
OM thanks Shunsaku Horiuchi for helpful discussions.  This work made use of computing resources and support provided by the Physics Institute at the University of Antioquia. OM was partially supported by COLCIENCIAS through the grant number 111-556-934918. This research has been supported at IAP by the ERC Project No.~267117 (DARK) hosted by Universit\'e Pierre et Marie Curie (UPMC) - Paris 6 and at JHU by NSF Grant No.~OIA-1124403. This work has been also supported by UPMC and STFC. Finally, this project has been carried out in the ILP LABEX (ANR-10-LABX-63) and has been supported by French state funds managed by the ANR, within the Investissements d'Avenir programme (ANR-11-IDEX-0004-02).
\end{acknowledgements}

\vspace{-0.3cm}

\appendix
\section{Emission and loss terms}

\subsection{Emission}

The IC emission spectrum reads (see e.g.~Refs.~\cite{Blumenthal1970,Cirelli2009,Cirelli_cookbook})
\begin{align}
\label{P_IC}
P_{\mathrm{IC}}(\vec{x},E_{\gamma},E_{\mathrm{e}}) = \dfrac{3 \sigma_{\mathrm{T}} c}{4 \gamma_{\mathrm{L}}^{2}} \int_{1/4\gamma_{\mathrm{L}}^{2}}^{1} \! \mathrm{d}q \left( E_{\gamma} - E_{\gamma}^{0}(q) \right) \dfrac{n(\vec{x},E_{\gamma}^{0}(q))}{q} \nonumber \\ 
\times \left( 2q\ln q + q + 1 -2q^{2} +\dfrac{1}{2} \dfrac{\epsilon^{2}}{1-\epsilon} (1-q)\right),
\end{align}
where $\sigma_{\mathrm{T}}$ is the Thomson cross-section, $c$ the speed of light, $\gamma_{\mathrm{L}}$ the Lorentz factor of the electrons, $\epsilon = E_{\gamma}/E_{\mathrm{e}}$ and the initial energy of the photon of the ISRF is related to variable $q$ via:
\begin{equation}
E_{\gamma}^{0}(q) = \dfrac{E_{\gamma}}{4 q \gamma_{\mathrm{L}}^{2} (1-\epsilon)}.
\end{equation}
In Eq.~\eqref{P_IC}, $n$ is the sum of the number densities per unit energy for the different components of the photon bath, namely cosmic microwave background (CMB) radiation, infrared (IR) radiation from dust and stellar photons: $n = n_{\mathrm{CMB}} + n_{\mathrm{IR}} + n_{\mathrm{stellar}}$. The corresponding maps are taken from Ref.~\cite{Cirelli_cookbook_secondaries}.

For bremsstrahlung, the emission term is the sum of the contributions from neutral and ionized gas, and reads \cite{Cirelli_cookbook_secondaries}: 
\begin{align}
P_{\mathrm{brems}}(\vec{x},E_{\gamma},E_{\mathrm{e}}) = c E_{\gamma} & \left[ (n_{\mathrm{HI}}(\vec{x}) + 2 n_{\mathrm{H_{2}}}(\vec{x}))  \dfrac{\mathrm{d}\sigma_{\mathrm{H}}}{\mathrm{d}E_{\gamma}}(E_{\gamma},E_{\mathrm{e}}) \right. \nonumber \\
& \left. + n_{\mathrm{He}}(\vec{x}) \dfrac{\mathrm{d}\sigma_{\mathrm{He}}}{\mathrm{d}E_{\gamma}}(E_{\gamma},E_{\mathrm{e}}) \right. \nonumber \\
& \left. + n_{\mathrm{HII}}(\vec{x}) \dfrac{\mathrm{d}\sigma_{\mathrm{HII}}}{\mathrm{d}E_{\gamma}}(E_{\gamma},E_{\mathrm{e}})\right] ,
\end{align}
with the differential cross section given by
\begin{equation}
\dfrac{\mathrm{d}\sigma_{a}}{\mathrm{d}E_{\gamma}} = \dfrac{3 \alpha \sigma_{\mathrm{T}}}{8 \pi E_{\gamma}} \left[ \left( 1 + \left( 1 - \dfrac{E_{\gamma}}{E_{\mathrm{e}}} \right) ^{2}\right) \phi_{1,a} - \dfrac{2}{3} \left( 1 - \dfrac{E_{\gamma}}{E_{\mathrm{e}}} \right) \phi_{2,a}\right]
\end{equation}
where $\alpha$ is the fine structure constant. $a = $ H, He or HII, with $\phi_{1,\mathrm{H}} = 45.79$, $\phi_{2,\mathrm{H}} = 44.46$, $\phi_{1,\mathrm{He}} = 134.6$, $\phi_{2,\mathrm{He}} = 131.4$ and for ionized hydrogen:
\begin{align}
\phi_{1,\mathrm{HII}}(E_{\gamma},E_{\mathrm{e}}) & = \phi_{2,\mathrm{HII}}(E_{\gamma},E_{\mathrm{e}}) \nonumber \\
& = 4 Z (Z + 1) \left[ \ln \left( \dfrac{2E_{\mathrm{e}}}{m_{\mathrm{e}}c^{2}} \left( \dfrac{E_{\mathrm{e}} - E_{\gamma}}{E_{\gamma}} \right) \right) - \dfrac{1}{2} \right],
\end{align}
with Z = 1 for hydrogen and $m_{\mathrm{e}}$ the electron mass.

\subsection{Losses}

The total energy loss term is the sum of the synchrotron, IC and bremsstrahlung contributions, $b = b_{\mathrm{syn}} + b_{\mathrm{IC}} + b_{\mathrm{brems}}$. Ionization and Coulomb losses are negligible at the energies of interest and for the GC region.

The synchrotron loss term reads (see e.g.~Ref.~\cite{Cirelli_cookbook_secondaries})
\begin{equation}
b_{\mathrm{syn}}(\vec{x},E) = \dfrac{4}{3} \sigma_{\mathrm{T}} c \dfrac{B(\vec{x})^{2}}{2 \mu_{0}} \gamma_{\mathrm{L}}, 
\end{equation}
where $\mu_{0}$ is the vacuum permeability. For the magnetic field $B$, we consider Model 1 of Ref.~\cite{Cirelli_cookbook_secondaries} which reads, in cylindrical coordinates:
\begin{equation}
B(\vec{x}) = B(r,z) = B_{0} \exp \left( - \dfrac{r-r_{\odot}}{r_{\mathrm{B}}} - \dfrac{|z|}{z_{\mathrm{B}}} \right),
\end{equation}
with $B_{0} \approx 5\ \mu \rm G$, $r_{\odot} = 8.25\ \rm kpc$, $r_{\mathrm{B}} = 10\ \rm kpc$ and $z_{\mathrm{B}} = 2\ \rm kpc$.

The bremsstrahlung loss term is given by the integral of the emission term:
\begin{equation}
b_{\mathrm{brems}}(\vec{x},E) = \int_{0}^{E} \! P_{\mathrm{brems}}(\vec{x},E_{\gamma},E) \, \mathrm{d}E_{\gamma}.
\end{equation}
It is the sum of the contributions from ionized and neutral gas, $b_{\mathrm{brems}} = b_{\mathrm{brems,I}} + b_{\mathrm{brems,N}}$, where:
\begin{equation}
b_{\mathrm{brems,I}}(\vec{x},E) = \alpha \dfrac{3 \sigma_{\mathrm{T}}c}{2\pi} n_{\mathrm{HII}}(\vec{x}) Z(Z+1) \left( \ln \left( \dfrac{2E}{m_{\mathrm{e}}}\right) -\dfrac{1}{3} \right) E,
\end{equation}
\begin{multline}
b_{\mathrm{brems,N}}(\vec{x},E)\\ 
 \quad\quad= \alpha \dfrac{3 \sigma_{\mathrm{T}}c}{8\pi} E \left[ \left( n_{\mathrm{HI}}(\vec{x}) + 2 n_{\mathrm{H_{2}}}(\vec{x}) \right) 
\left( \dfrac{4}{3} \phi_{1,\mathrm{H}} -\dfrac{1}{3} \phi_{2,\mathrm{H}} \right) \right. \\
\left. + n_{\mathrm{He}}(\vec{x}) \left( \dfrac{4}{3} \phi_{1,\mathrm{He}} -\dfrac{1}{3} \phi_{2,\mathrm{He}} \right) \right],
\end{multline}
with $Z = 1$ for hydrogen. The gas density maps for $n_{\mathrm{HI}}$, $n_{\mathrm{HII}}$, $n_{\mathrm{He}}$ and $n_{\mathrm{H_{2}}}$ are taken from Ref.~\cite{Cirelli_cookbook_secondaries}.

Similarly, the IC loss term is given by
\begin{equation}
b_{\mathrm{IC}}(\vec{x},E) = \int_{0}^{E} \! P_{\mathrm{IC}}(\vec{x},E_{\gamma},E) \, \mathrm{d}E_{\gamma}.
\end{equation}
and tabulated for convenience. 
\section{Comparison with DRAGON}
\label{DRAGONCompare}
As discussed in Sec.~\ref{sec:secondaryemission} our approach to secondaries is computationally more straight forward than using codes such as \textsc{Dragon} or \textsc{Galprop}. In Fig.~\ref{fig:DRAGONcomparison} we compare our results to published results from \textsc{Dragon}. The differences seen in the bremsstrahlung results, at high latitude, are not important as the order of magnitude is similar and in the cases we consider bremsstrahlung has a negligible contribution. Accounting for the uncertainties in the diffusion coefficient, ISRF and other relevant parameters, our IC results are a reasonable approximation to those found in Ref.~\cite{Cirelli_et_al_constraints}. Therefore, using \textsc{Dragon}, instead of our derivation of secondaries, would not significantly change the conclusions of our article.

\bibliography{3D}

\begin{thebibliography}{66}%
\makeatletter
\providecommand \@ifxundefined [1]{%
 \@ifx{#1\undefined}
}%
\providecommand \@ifnum [1]{%
 \ifnum #1\expandafter \@firstoftwo
 \else \expandafter \@secondoftwo
 \fi
}%
\providecommand \@ifx [1]{%
 \ifx #1\expandafter \@firstoftwo
 \else \expandafter \@secondoftwo
 \fi
}%
\providecommand \natexlab [1]{#1}%
\providecommand \enquote  [1]{``#1''}%
\providecommand \bibnamefont  [1]{#1}%
\providecommand \bibfnamefont [1]{#1}%
\providecommand \citenamefont [1]{#1}%
\providecommand \href@noop [0]{\@secondoftwo}%
\providecommand \href [0]{\begingroup \@sanitize@url \@href}%
\providecommand \@href[1]{\@@startlink{#1}\@@href}%
\providecommand \@@href[1]{\endgroup#1\@@endlink}%
\providecommand \@sanitize@url [0]{\catcode `\\12\catcode `\$12\catcode
  `\&12\catcode `\#12\catcode `\^12\catcode `\_12\catcode `\%12\relax}%
\providecommand \@@startlink[1]{}%
\providecommand \@@endlink[0]{}%
\providecommand \url  [0]{\begingroup\@sanitize@url \@url }%
\providecommand \@url [1]{\endgroup\@href {#1}{\urlprefix }}%
\providecommand \urlprefix  [0]{URL }%
\providecommand \Eprint [0]{\href }%
\providecommand \doibase [0]{http://dx.doi.org/}%
\providecommand \selectlanguage [0]{\@gobble}%
\providecommand \bibinfo  [0]{\@secondoftwo}%
\providecommand \bibfield  [0]{\@secondoftwo}%
\providecommand \translation [1]{[#1]}%
\providecommand \BibitemOpen [0]{}%
\providecommand \bibitemStop [0]{}%
\providecommand \bibitemNoStop [0]{.\EOS\space}%
\providecommand \EOS [0]{\spacefactor3000\relax}%
\providecommand \BibitemShut  [1]{\csname bibitem#1\endcsname}%
\let\auto@bib@innerbib\@empty
\bibitem [{\citenamefont {{Goodenough}}\ and\ \citenamefont
  {{Hooper}}(2009)}]{Goodenough2009gk}%
  \BibitemOpen
  \bibfield  {author} {\bibinfo {author} {\bibfnamefont {L.}~\bibnamefont
  {{Goodenough}}}\ and\ \bibinfo {author} {\bibfnamefont {D.}~\bibnamefont
  {{Hooper}}},\ }\href@noop {} {\bibfield  {journal} {\bibinfo  {journal}
  {ArXiv e-prints}\ } (\bibinfo {year} {2009})},\ \Eprint
  {http://arxiv.org/abs/0910.2998} {arXiv:0910.2998 [hep-ph]} \BibitemShut
  {NoStop}%
\bibitem [{\citenamefont {{Vitale}}\ \emph {et~al.}(2009)\citenamefont
  {{Vitale}}, \citenamefont {{Morselli}},\ and\ \citenamefont {{for the
  Fermi/LAT Collaboration}}}]{Vitale:2009hr}%
  \BibitemOpen
  \bibfield  {author} {\bibinfo {author} {\bibfnamefont {V.}~\bibnamefont
  {{Vitale}}}, \bibinfo {author} {\bibfnamefont {A.}~\bibnamefont
  {{Morselli}}}, \ and\ \bibinfo {author} {\bibnamefont {{for the Fermi/LAT
  Collaboration}}},\ }\href@noop {} {\bibfield  {journal} {\bibinfo  {journal}
  {ArXiv e-prints}\ } (\bibinfo {year} {2009})},\ \Eprint
  {http://arxiv.org/abs/0912.3828} {arXiv:0912.3828 [astro-ph.HE]} \BibitemShut
  {NoStop}%
\bibitem [{\citenamefont {Hooper}\ and\ \citenamefont
  {Goodenough}(2011)}]{Hooper:2010mq}%
  \BibitemOpen
  \bibfield  {author} {\bibinfo {author} {\bibfnamefont {D.}~\bibnamefont
  {Hooper}}\ and\ \bibinfo {author} {\bibfnamefont {L.}~\bibnamefont
  {Goodenough}},\ }\href {\doibase 10.1016/j.physletb.2011.02.029} {\bibfield
  {journal} {\bibinfo  {journal} {Phys.Lett.}\ }\textbf {\bibinfo {volume}
  {B697}},\ \bibinfo {pages} {412} (\bibinfo {year} {2011})},\ \Eprint
  {http://arxiv.org/abs/1010.2752} {arXiv:1010.2752 [hep-ph]} \BibitemShut
  {NoStop}%
\bibitem [{\citenamefont {Hooper}\ and\ \citenamefont {Linden}(2011)}]{hooper}%
  \BibitemOpen
  \bibfield  {author} {\bibinfo {author} {\bibfnamefont {D.}~\bibnamefont
  {Hooper}}\ and\ \bibinfo {author} {\bibfnamefont {T.}~\bibnamefont
  {Linden}},\ }\href {\doibase 10.1103/PhysRevD.84.123005} {\bibfield
  {journal} {\bibinfo  {journal} {Phys.Rev.}\ }\textbf {\bibinfo {volume}
  {D84}},\ \bibinfo {pages} {123005} (\bibinfo {year} {2011})},\ \Eprint
  {http://arxiv.org/abs/1110.0006} {arXiv:1110.0006 [astro-ph.HE]} \BibitemShut
  {NoStop}%
\bibitem [{\citenamefont {Abazajian}\ and\ \citenamefont
  {Kaplinghat}(2012)}]{AbazajianKaplinghat2012}%
  \BibitemOpen
  \bibfield  {author} {\bibinfo {author} {\bibfnamefont {K.~N.}\ \bibnamefont
  {Abazajian}}\ and\ \bibinfo {author} {\bibfnamefont {M.}~\bibnamefont
  {Kaplinghat}},\ }\href {\doibase 10.1103/PhysRevD.86.083511} {\bibfield
  {journal} {\bibinfo  {journal} {Phys.Rev.}\ }\textbf {\bibinfo {volume}
  {D86}},\ \bibinfo {pages} {083511} (\bibinfo {year} {2012})},\ \Eprint
  {http://arxiv.org/abs/1207.6047} {arXiv:1207.6047 [astro-ph.HE]} \BibitemShut
  {NoStop}%
\bibitem [{\citenamefont {{Abazajian}}\ and\ \citenamefont
  {{Kaplinghat}}(2013)}]{AbazajianKaplinghat2013}%
  \BibitemOpen
  \bibfield  {author} {\bibinfo {author} {\bibfnamefont {K.~N.}\ \bibnamefont
  {{Abazajian}}}\ and\ \bibinfo {author} {\bibfnamefont {M.}~\bibnamefont
  {{Kaplinghat}}},\ }\href {\doibase 10.1103/PhysRevD.87.129902} {\bibfield
  {journal} {\bibinfo  {journal} {\prd}\ }\textbf {\bibinfo {volume} {87}},\
  \bibinfo {eid} {129902} (\bibinfo {year} {2013})}\BibitemShut {NoStop}%
\bibitem [{\citenamefont {Gordon}\ and\ \citenamefont
  {Macias}(2013)}]{GordonMacias2013}%
  \BibitemOpen
  \bibfield  {author} {\bibinfo {author} {\bibfnamefont {C.}~\bibnamefont
  {Gordon}}\ and\ \bibinfo {author} {\bibfnamefont {O.}~\bibnamefont
  {Macias}},\ }\href {\doibase 10.1103/PhysRevD.88.083521} {\bibfield
  {journal} {\bibinfo  {journal} {Phys.Rev.}\ }\textbf {\bibinfo {volume}
  {D88}},\ \bibinfo {pages} {083521} (\bibinfo {year} {2013})},\ \Eprint
  {http://arxiv.org/abs/1306.5725} {arXiv:1306.5725 [astro-ph.HE]} \BibitemShut
  {NoStop}%
\bibitem [{\citenamefont {{The Fermi-LAT
  Collaboration}}(2015)}]{FermiGalacticCenter2015}%
  \BibitemOpen
  \bibfield  {author} {\bibinfo {author} {\bibnamefont {{The Fermi-LAT
  Collaboration}}},\ }\href@noop {} {\bibfield  {journal} {\bibinfo  {journal}
  {ArXiv e-prints}\ } (\bibinfo {year} {2015})},\ \Eprint
  {http://arxiv.org/abs/1511.02938} {arXiv:1511.02938 [astro-ph.HE]}
  \BibitemShut {NoStop}%
\bibitem [{\citenamefont {{Hooper}}\ and\ \citenamefont
  {{Slatyer}}(2013)}]{hooperslatyer2013}%
  \BibitemOpen
  \bibfield  {author} {\bibinfo {author} {\bibfnamefont {D.}~\bibnamefont
  {{Hooper}}}\ and\ \bibinfo {author} {\bibfnamefont {T.~R.}\ \bibnamefont
  {{Slatyer}}},\ }\href {\doibase 10.1016/j.dark.2013.06.003} {\bibfield
  {journal} {\bibinfo  {journal} {Physics of the Dark Universe}\ }\textbf
  {\bibinfo {volume} {2}},\ \bibinfo {pages} {118} (\bibinfo {year} {2013})},\
  \Eprint {http://arxiv.org/abs/1302.6589} {arXiv:1302.6589 [astro-ph.HE]}
  \BibitemShut {NoStop}%
\bibitem [{\citenamefont {Daylan}\ \emph {et~al.}(2014)\citenamefont {Daylan},
  \citenamefont {Finkbeiner}, \citenamefont {Hooper}, \citenamefont {Linden},
  \citenamefont {Portillo} \emph {et~al.}}]{Daylan:2014}%
  \BibitemOpen
  \bibfield  {author} {\bibinfo {author} {\bibfnamefont {T.}~\bibnamefont
  {Daylan}}, \bibinfo {author} {\bibfnamefont {D.~P.}\ \bibnamefont
  {Finkbeiner}}, \bibinfo {author} {\bibfnamefont {D.}~\bibnamefont {Hooper}},
  \bibinfo {author} {\bibfnamefont {T.}~\bibnamefont {Linden}}, \bibinfo
  {author} {\bibfnamefont {S.~K.~N.}\ \bibnamefont {Portillo}},  \emph
  {et~al.},\ }\href@noop {} {\bibfield  {journal} {\bibinfo  {journal} {ArXiv
  e-prints}\ } (\bibinfo {year} {2014})},\ \Eprint
  {http://arxiv.org/abs/1402.6703} {arXiv:1402.6703 [astro-ph.HE]} \BibitemShut
  {NoStop}%
\bibitem [{\citenamefont {Macias}\ and\ \citenamefont
  {Gordon}(2014)}]{MaciasGordon2014}%
  \BibitemOpen
  \bibfield  {author} {\bibinfo {author} {\bibfnamefont {O.}~\bibnamefont
  {Macias}}\ and\ \bibinfo {author} {\bibfnamefont {C.}~\bibnamefont
  {Gordon}},\ }\href@noop {} {\bibfield  {journal} {\bibinfo  {journal}
  {Phys.Rev.}\ }\textbf {\bibinfo {volume} {D}} (\bibinfo {year} {2014})},\
  \Eprint {http://arxiv.org/abs/1312.6671} {arXiv:1312.6671 [astro-ph.HE]}
  \BibitemShut {NoStop}%
\bibitem [{\citenamefont {Abazajian}\ \emph {et~al.}(2014)\citenamefont
  {Abazajian}, \citenamefont {Canac}, \citenamefont {Horiuchi},\ and\
  \citenamefont {Kaplinghat}}]{Abazajian2014}%
  \BibitemOpen
  \bibfield  {author} {\bibinfo {author} {\bibfnamefont {K.~N.}\ \bibnamefont
  {Abazajian}}, \bibinfo {author} {\bibfnamefont {N.}~\bibnamefont {Canac}},
  \bibinfo {author} {\bibfnamefont {S.}~\bibnamefont {Horiuchi}}, \ and\
  \bibinfo {author} {\bibfnamefont {M.}~\bibnamefont {Kaplinghat}},\ }\href
  {\doibase 10.1103/PhysRevD.90.023526} {\bibfield  {journal} {\bibinfo
  {journal} {Phys. Rev.}\ }\textbf {\bibinfo {volume} {D90}},\ \bibinfo {pages}
  {023526} (\bibinfo {year} {2014})},\ \Eprint {http://arxiv.org/abs/1402.4090}
  {arXiv:1402.4090 [astro-ph.HE]} \BibitemShut {NoStop}%
\bibitem [{\citenamefont {{Zhou}}\ \emph {et~al.}(2015)\citenamefont {{Zhou}},
  \citenamefont {{Liang}}, \citenamefont {{Huang}}, \citenamefont {{Li}},
  \citenamefont {{Fan}}, \citenamefont {{Feng}},\ and\ \citenamefont
  {{Chang}}}]{Zho2015}%
  \BibitemOpen
  \bibfield  {author} {\bibinfo {author} {\bibfnamefont {B.}~\bibnamefont
  {{Zhou}}}, \bibinfo {author} {\bibfnamefont {Y.-F.}\ \bibnamefont {{Liang}}},
  \bibinfo {author} {\bibfnamefont {X.}~\bibnamefont {{Huang}}}, \bibinfo
  {author} {\bibfnamefont {X.}~\bibnamefont {{Li}}}, \bibinfo {author}
  {\bibfnamefont {Y.-Z.}\ \bibnamefont {{Fan}}}, \bibinfo {author}
  {\bibfnamefont {L.}~\bibnamefont {{Feng}}}, \ and\ \bibinfo {author}
  {\bibfnamefont {J.}~\bibnamefont {{Chang}}},\ }\href {\doibase
  10.1103/PhysRevD.91.123010} {\bibfield  {journal} {\bibinfo  {journal}
  {\prd}\ }\textbf {\bibinfo {volume} {91}},\ \bibinfo {eid} {123010} (\bibinfo
  {year} {2015})},\ \Eprint {http://arxiv.org/abs/1406.6948} {arXiv:1406.6948
  [astro-ph.HE]} \BibitemShut {NoStop}%
\bibitem [{\citenamefont {{Calore}}\ \emph {et~al.}(2015)\citenamefont
  {{Calore}}, \citenamefont {{Cholis}},\ and\ \citenamefont
  {{Weniger}}}]{CaloreCholisWeniger2015}%
  \BibitemOpen
  \bibfield  {author} {\bibinfo {author} {\bibfnamefont {F.}~\bibnamefont
  {{Calore}}}, \bibinfo {author} {\bibfnamefont {I.}~\bibnamefont {{Cholis}}},
  \ and\ \bibinfo {author} {\bibfnamefont {C.}~\bibnamefont {{Weniger}}},\
  }\href {\doibase 10.1088/1475-7516/2015/03/038} {\bibfield  {journal}
  {\bibinfo  {journal} {JCAP}\ }\textbf {\bibinfo {volume} {3}},\ \bibinfo
  {eid} {038} (\bibinfo {year} {2015})},\ \Eprint
  {http://arxiv.org/abs/1409.0042} {arXiv:1409.0042} \BibitemShut {NoStop}%
\bibitem [{\citenamefont {{de Boer}}\ \emph {et~al.}(2015)\citenamefont {{de
  Boer}}, \citenamefont {{Gebauer}}, \citenamefont {{Kunz}},\ and\
  \citenamefont {{Neumann}}}]{deBoer2015}%
  \BibitemOpen
  \bibfield  {author} {\bibinfo {author} {\bibfnamefont {W.}~\bibnamefont {{de
  Boer}}}, \bibinfo {author} {\bibfnamefont {I.}~\bibnamefont {{Gebauer}}},
  \bibinfo {author} {\bibfnamefont {S.}~\bibnamefont {{Kunz}}}, \ and\ \bibinfo
  {author} {\bibfnamefont {A.}~\bibnamefont {{Neumann}}},\ }\href@noop {}
  {\bibfield  {journal} {\bibinfo  {journal} {ArXiv e-prints}\ } (\bibinfo
  {year} {2015})},\ \Eprint {http://arxiv.org/abs/1509.05310} {arXiv:1509.05310
  [astro-ph.HE]} \BibitemShut {NoStop}%
\bibitem [{\citenamefont {Gaggero}\ \emph {et~al.}(2015)\citenamefont
  {Gaggero}, \citenamefont {Taoso}, \citenamefont {Urbano}, \citenamefont
  {Valli},\ and\ \citenamefont {Ullio}}]{Gaggero2015}%
  \BibitemOpen
  \bibfield  {author} {\bibinfo {author} {\bibfnamefont {D.}~\bibnamefont
  {Gaggero}}, \bibinfo {author} {\bibfnamefont {M.}~\bibnamefont {Taoso}},
  \bibinfo {author} {\bibfnamefont {A.}~\bibnamefont {Urbano}}, \bibinfo
  {author} {\bibfnamefont {M.}~\bibnamefont {Valli}}, \ and\ \bibinfo {author}
  {\bibfnamefont {P.}~\bibnamefont {Ullio}},\ }\href@noop {} {\  (\bibinfo
  {year} {2015})},\ \Eprint {http://arxiv.org/abs/1507.06129} {arXiv:1507.06129
  [astro-ph.HE]} \BibitemShut {NoStop}%
\bibitem [{\citenamefont {{Carlson}}\ \emph {et~al.}(2015)\citenamefont
  {{Carlson}}, \citenamefont {{Linden}},\ and\ \citenamefont
  {{Profumo}}}]{CarlsonLindenProfumo2015}%
  \BibitemOpen
  \bibfield  {author} {\bibinfo {author} {\bibfnamefont {E.}~\bibnamefont
  {{Carlson}}}, \bibinfo {author} {\bibfnamefont {T.}~\bibnamefont {{Linden}}},
  \ and\ \bibinfo {author} {\bibfnamefont {S.}~\bibnamefont {{Profumo}}},\
  }\href@noop {} {\bibfield  {journal} {\bibinfo  {journal} {ArXiv e-prints}\ }
  (\bibinfo {year} {2015})},\ \Eprint {http://arxiv.org/abs/1510.04698}
  {arXiv:1510.04698 [astro-ph.HE]} \BibitemShut {NoStop}%
\bibitem [{\citenamefont {Carlson}\ \emph {et~al.}(2016)\citenamefont
  {Carlson}, \citenamefont {Linden},\ and\ \citenamefont
  {Profumo}}]{CarlsonLindenProfumo2016}%
  \BibitemOpen
  \bibfield  {author} {\bibinfo {author} {\bibfnamefont {E.}~\bibnamefont
  {Carlson}}, \bibinfo {author} {\bibfnamefont {T.}~\bibnamefont {Linden}}, \
  and\ \bibinfo {author} {\bibfnamefont {S.}~\bibnamefont {Profumo}},\
  }\href@noop {} {\  (\bibinfo {year} {2016})},\ \Eprint
  {http://arxiv.org/abs/1603.06584} {arXiv:1603.06584 [astro-ph.HE]}
  \BibitemShut {NoStop}%
\bibitem [{\citenamefont {Aharonian}\ \emph {et~al.}(2006)\citenamefont
  {Aharonian} \emph {et~al.}}]{Aharonian:2006}%
  \BibitemOpen
  \bibfield  {author} {\bibinfo {author} {\bibfnamefont {F.}~\bibnamefont
  {Aharonian}} \emph {et~al.} (\bibinfo {collaboration} {H.E.S.S.}),\ }\href
  {\doibase 10.1038/nature04467} {\bibfield  {journal} {\bibinfo  {journal}
  {Nature}\ }\textbf {\bibinfo {volume} {439}},\ \bibinfo {pages} {695}
  (\bibinfo {year} {2006})},\ \Eprint {http://arxiv.org/abs/astro-ph/0603021}
  {arXiv:astro-ph/0603021 [astro-ph]} \BibitemShut {NoStop}%
\bibitem [{\citenamefont {Yusef-Zadeh}\ \emph {et~al.}(2013)\citenamefont
  {Yusef-Zadeh}, \citenamefont {Hewitt}, \citenamefont {Wardle}, \citenamefont
  {Tatischeff}, \citenamefont {Roberts} \emph {et~al.}}]{Yusef-Zadeh2013}%
  \BibitemOpen
  \bibfield  {author} {\bibinfo {author} {\bibfnamefont {F.}~\bibnamefont
  {Yusef-Zadeh}}, \bibinfo {author} {\bibfnamefont {J.}~\bibnamefont {Hewitt}},
  \bibinfo {author} {\bibfnamefont {M.}~\bibnamefont {Wardle}}, \bibinfo
  {author} {\bibfnamefont {V.}~\bibnamefont {Tatischeff}}, \bibinfo {author}
  {\bibfnamefont {D.}~\bibnamefont {Roberts}},  \emph {et~al.},\ }\href
  {\doibase 10.1088/0004-637X/762/1/33} {\bibfield  {journal} {\bibinfo
  {journal} {Astrophys.J.}\ }\textbf {\bibinfo {volume} {762}},\ \bibinfo
  {pages} {33} (\bibinfo {year} {2013})},\ \Eprint
  {http://arxiv.org/abs/1206.6882} {arXiv:1206.6882 [astro-ph.HE]} \BibitemShut
  {NoStop}%
\bibitem [{\citenamefont {{Yoast-Hull}}\ \emph {et~al.}(2014)\citenamefont
  {{Yoast-Hull}}, \citenamefont {{Gallagher}},\ and\ \citenamefont
  {{Zweibel}}}]{YoastHullGallagherZweibel2014}%
  \BibitemOpen
  \bibfield  {author} {\bibinfo {author} {\bibfnamefont {T.~M.}\ \bibnamefont
  {{Yoast-Hull}}}, \bibinfo {author} {\bibfnamefont {J.~S.}\ \bibnamefont
  {{Gallagher}}, \bibfnamefont {III}}, \ and\ \bibinfo {author} {\bibfnamefont
  {E.~G.}\ \bibnamefont {{Zweibel}}},\ }\href {\doibase
  10.1088/0004-637X/790/2/86} {\bibfield  {journal} {\bibinfo  {journal}
  {\apj}\ }\textbf {\bibinfo {volume} {790}},\ \bibinfo {eid} {86} (\bibinfo
  {year} {2014})},\ \Eprint {http://arxiv.org/abs/1405.7059} {arXiv:1405.7059
  [astro-ph.HE]} \BibitemShut {NoStop}%
\bibitem [{\citenamefont {{Macias}}\ \emph {et~al.}(2015)\citenamefont
  {{Macias}}, \citenamefont {{Gordon}}, \citenamefont {{Crocker}},\ and\
  \citenamefont {{Profumo}}}]{MacGorCroProf2015}%
  \BibitemOpen
  \bibfield  {author} {\bibinfo {author} {\bibfnamefont {O.}~\bibnamefont
  {{Macias}}}, \bibinfo {author} {\bibfnamefont {C.}~\bibnamefont {{Gordon}}},
  \bibinfo {author} {\bibfnamefont {R.~M.}\ \bibnamefont {{Crocker}}}, \ and\
  \bibinfo {author} {\bibfnamefont {S.}~\bibnamefont {{Profumo}}},\ }\href
  {\doibase 10.1093/mnras/stv1002} {\bibfield  {journal} {\bibinfo  {journal}
  {\mnras}\ }\textbf {\bibinfo {volume} {451}},\ \bibinfo {pages} {1833}
  (\bibinfo {year} {2015})},\ \Eprint {http://arxiv.org/abs/1410.1678}
  {arXiv:1410.1678 [astro-ph.HE]} \BibitemShut {NoStop}%
\bibitem [{\citenamefont {Abazajian}(2011)}]{Abazajian:2010zy}%
  \BibitemOpen
  \bibfield  {author} {\bibinfo {author} {\bibfnamefont {K.~N.}\ \bibnamefont
  {Abazajian}},\ }\href {\doibase 10.1088/1475-7516/2011/03/010} {\bibfield
  {journal} {\bibinfo  {journal} {JCAP}\ }\textbf {\bibinfo {volume} {1103}},\
  \bibinfo {pages} {010} (\bibinfo {year} {2011})},\ \Eprint
  {http://arxiv.org/abs/1011.4275} {arXiv:1011.4275 [astro-ph.HE]} \BibitemShut
  {NoStop}%
\bibitem [{\citenamefont {Wharton}\ \emph {et~al.}(2012)\citenamefont
  {Wharton}, \citenamefont {Chatterjee}, \citenamefont {Cordes}, \citenamefont
  {Deneva},\ and\ \citenamefont {Lazio}}]{Wharton2012}%
  \BibitemOpen
  \bibfield  {author} {\bibinfo {author} {\bibfnamefont {R.~S.}\ \bibnamefont
  {Wharton}}, \bibinfo {author} {\bibfnamefont {S.}~\bibnamefont {Chatterjee}},
  \bibinfo {author} {\bibfnamefont {J.~M.}\ \bibnamefont {Cordes}}, \bibinfo
  {author} {\bibfnamefont {J.~S.}\ \bibnamefont {Deneva}}, \ and\ \bibinfo
  {author} {\bibfnamefont {T.~J.~W.}\ \bibnamefont {Lazio}},\ }\href
  {http://stacks.iop.org/0004-637X/753/i=2/a=108} {\bibfield  {journal}
  {\bibinfo  {journal} {The Astrophysical Journal}\ }\textbf {\bibinfo {volume}
  {753}},\ \bibinfo {pages} {108} (\bibinfo {year} {2012})}\BibitemShut
  {NoStop}%
\bibitem [{\citenamefont {{Gordon}}\ and\ \citenamefont
  {{Mac{\'{\i}}as}}(2014)}]{GordonMacias2013erratum}%
  \BibitemOpen
  \bibfield  {author} {\bibinfo {author} {\bibfnamefont {C.}~\bibnamefont
  {{Gordon}}}\ and\ \bibinfo {author} {\bibfnamefont {O.}~\bibnamefont
  {{Mac{\'{\i}}as}}},\ }\href {\doibase 10.1103/PhysRevD.89.049901} {\bibfield
  {journal} {\bibinfo  {journal} {\prd}\ }\textbf {\bibinfo {volume} {89}},\
  \bibinfo {eid} {049901} (\bibinfo {year} {2014})}\BibitemShut {NoStop}%
\bibitem [{\citenamefont {{Mirabal}}(2013)}]{Mirabal2013}%
  \BibitemOpen
  \bibfield  {author} {\bibinfo {author} {\bibfnamefont {N.}~\bibnamefont
  {{Mirabal}}},\ }\href {\doibase 10.1093/mnras/stt1740} {\bibfield  {journal}
  {\bibinfo  {journal} {\mnras}\ }\textbf {\bibinfo {volume} {436}},\ \bibinfo
  {pages} {2461} (\bibinfo {year} {2013})},\ \Eprint
  {http://arxiv.org/abs/1309.3428} {arXiv:1309.3428 [astro-ph.HE]} \BibitemShut
  {NoStop}%
\bibitem [{\citenamefont {{Yuan}}\ and\ \citenamefont
  {{Zhang}}(2014)}]{YuanZhang2014}%
  \BibitemOpen
  \bibfield  {author} {\bibinfo {author} {\bibfnamefont {Q.}~\bibnamefont
  {{Yuan}}}\ and\ \bibinfo {author} {\bibfnamefont {B.}~\bibnamefont
  {{Zhang}}},\ }\href {\doibase 10.1016/j.jheap.2014.06.001} {\bibfield
  {journal} {\bibinfo  {journal} {Journal of High Energy Astrophysics}\
  }\textbf {\bibinfo {volume} {3}},\ \bibinfo {pages} {1} (\bibinfo {year}
  {2014})},\ \Eprint {http://arxiv.org/abs/1404.2318} {arXiv:1404.2318
  [astro-ph.HE]} \BibitemShut {NoStop}%
\bibitem [{\citenamefont {{Brandt}}\ and\ \citenamefont
  {{Kocsis}}(2015)}]{BrandtKocsis2015}%
  \BibitemOpen
  \bibfield  {author} {\bibinfo {author} {\bibfnamefont {T.~D.}\ \bibnamefont
  {{Brandt}}}\ and\ \bibinfo {author} {\bibfnamefont {B.}~\bibnamefont
  {{Kocsis}}},\ }\href@noop {} {\bibfield  {journal} {\bibinfo  {journal}
  {ArXiv e-prints}\ } (\bibinfo {year} {2015})},\ \Eprint
  {http://arxiv.org/abs/1507.05616} {arXiv:1507.05616 [astro-ph.HE]}
  \BibitemShut {NoStop}%
\bibitem [{\citenamefont {{O'Leary}}\ \emph {et~al.}(2015)\citenamefont
  {{O'Leary}}, \citenamefont {{Kistler}}, \citenamefont {{Kerr}},\ and\
  \citenamefont {{Dexter}}}]{OLeary2015}%
  \BibitemOpen
  \bibfield  {author} {\bibinfo {author} {\bibfnamefont {R.~M.}\ \bibnamefont
  {{O'Leary}}}, \bibinfo {author} {\bibfnamefont {M.~D.}\ \bibnamefont
  {{Kistler}}}, \bibinfo {author} {\bibfnamefont {M.}~\bibnamefont {{Kerr}}}, \
  and\ \bibinfo {author} {\bibfnamefont {J.}~\bibnamefont {{Dexter}}},\
  }\href@noop {} {\bibfield  {journal} {\bibinfo  {journal} {ArXiv e-prints}\ }
  (\bibinfo {year} {2015})},\ \Eprint {http://arxiv.org/abs/1504.02477}
  {arXiv:1504.02477 [astro-ph.HE]} \BibitemShut {NoStop}%
\bibitem [{\citenamefont {Hooper}\ \emph {et~al.}(2013)\citenamefont {Hooper},
  \citenamefont {Cholis}, \citenamefont {Linden}, \citenamefont
  {Siegal-Gaskins},\ and\ \citenamefont {Slatyer}}]{Hooper2013}%
  \BibitemOpen
  \bibfield  {author} {\bibinfo {author} {\bibfnamefont {D.}~\bibnamefont
  {Hooper}}, \bibinfo {author} {\bibfnamefont {I.}~\bibnamefont {Cholis}},
  \bibinfo {author} {\bibfnamefont {T.}~\bibnamefont {Linden}}, \bibinfo
  {author} {\bibfnamefont {J.~M.}\ \bibnamefont {Siegal-Gaskins}}, \ and\
  \bibinfo {author} {\bibfnamefont {T.~R.}\ \bibnamefont {Slatyer}},\ }\href
  {\doibase 10.1103/PhysRevD.88.083009} {\bibfield  {journal} {\bibinfo
  {journal} {Phys. Rev. D}\ }\textbf {\bibinfo {volume} {88}},\ \bibinfo
  {pages} {083009} (\bibinfo {year} {2013})},\ \Eprint
  {http://arxiv.org/abs/1305.0830} {1305.0830 [astro-ph.HE]} \BibitemShut
  {NoStop}%
\bibitem [{\citenamefont {{Cholis}}\ \emph {et~al.}(2015)\citenamefont
  {{Cholis}}, \citenamefont {{Hooper}},\ and\ \citenamefont
  {{Linden}}}]{CholisHooperLinden2015}%
  \BibitemOpen
  \bibfield  {author} {\bibinfo {author} {\bibfnamefont {I.}~\bibnamefont
  {{Cholis}}}, \bibinfo {author} {\bibfnamefont {D.}~\bibnamefont {{Hooper}}},
  \ and\ \bibinfo {author} {\bibfnamefont {T.}~\bibnamefont {{Linden}}},\
  }\href {\doibase 10.1088/1475-7516/2015/06/043} {\bibfield  {journal}
  {\bibinfo  {journal} {JCAP}\ }\textbf {\bibinfo {volume} {6}},\ \bibinfo
  {eid} {043} (\bibinfo {year} {2015})},\ \Eprint
  {http://arxiv.org/abs/1407.5625} {arXiv:1407.5625 [astro-ph.HE]} \BibitemShut
  {NoStop}%
\bibitem [{\citenamefont {{Petrovi{\'c}}}\ \emph {et~al.}(2015)\citenamefont
  {{Petrovi{\'c}}}, \citenamefont {{Serpico}},\ and\ \citenamefont
  {{Zaharijas}}}]{PetrovicSerpicoZaharijas2015}%
  \BibitemOpen
  \bibfield  {author} {\bibinfo {author} {\bibfnamefont {J.}~\bibnamefont
  {{Petrovi{\'c}}}}, \bibinfo {author} {\bibfnamefont {P.~D.}\ \bibnamefont
  {{Serpico}}}, \ and\ \bibinfo {author} {\bibfnamefont {G.}~\bibnamefont
  {{Zaharijas}}},\ }\href {\doibase 10.1088/1475-7516/2015/02/023} {\bibfield
  {journal} {\bibinfo  {journal} {JCAP}\ }\textbf {\bibinfo {volume} {2}},\
  \bibinfo {eid} {023} (\bibinfo {year} {2015})},\ \Eprint
  {http://arxiv.org/abs/1411.2980} {arXiv:1411.2980 [astro-ph.HE]} \BibitemShut
  {NoStop}%
\bibitem [{\citenamefont {Lee}\ \emph {et~al.}(2015)\citenamefont {Lee},
  \citenamefont {Lisanti}, \citenamefont {Safdi}, \citenamefont {Slatyer},\
  and\ \citenamefont {Xue}}]{Lee2015}%
  \BibitemOpen
  \bibfield  {author} {\bibinfo {author} {\bibfnamefont {S.~K.}\ \bibnamefont
  {Lee}}, \bibinfo {author} {\bibfnamefont {M.}~\bibnamefont {Lisanti}},
  \bibinfo {author} {\bibfnamefont {B.~R.}\ \bibnamefont {Safdi}}, \bibinfo
  {author} {\bibfnamefont {T.~R.}\ \bibnamefont {Slatyer}}, \ and\ \bibinfo
  {author} {\bibfnamefont {W.}~\bibnamefont {Xue}},\ }\href@noop {} {\bibfield
  {journal} {\bibinfo  {journal} {ArXiv e-prints}\ } (\bibinfo {year}
  {2015})},\ \Eprint {http://arxiv.org/abs/1506.05124} {arXiv:1506.05124
  [astro-ph.HE]} \BibitemShut {NoStop}%
\bibitem [{\citenamefont {{Bartels}}\ \emph {et~al.}(2015)\citenamefont
  {{Bartels}}, \citenamefont {{Krishnamurthy}},\ and\ \citenamefont
  {{Weniger}}}]{BartelsKrishnamurthyWeniger2015}%
  \BibitemOpen
  \bibfield  {author} {\bibinfo {author} {\bibfnamefont {R.}~\bibnamefont
  {{Bartels}}}, \bibinfo {author} {\bibfnamefont {S.}~\bibnamefont
  {{Krishnamurthy}}}, \ and\ \bibinfo {author} {\bibfnamefont {C.}~\bibnamefont
  {{Weniger}}},\ }\href@noop {} {\bibfield  {journal} {\bibinfo  {journal}
  {ArXiv e-prints}\ } (\bibinfo {year} {2015})},\ \Eprint
  {http://arxiv.org/abs/1506.05104} {arXiv:1506.05104 [astro-ph.HE]}
  \BibitemShut {NoStop}%
\bibitem [{\citenamefont {{Linden}}(2015)}]{Linden2015}%
  \BibitemOpen
  \bibfield  {author} {\bibinfo {author} {\bibfnamefont {T.}~\bibnamefont
  {{Linden}}},\ }\href@noop {} {\bibfield  {journal} {\bibinfo  {journal}
  {ArXiv e-prints}\ } (\bibinfo {year} {2015})},\ \Eprint
  {http://arxiv.org/abs/1509.02928} {arXiv:1509.02928 [astro-ph.HE]}
  \BibitemShut {NoStop}%
\bibitem [{\citenamefont {{Abdo}}\ \emph {et~al.}(2013)\citenamefont {{Abdo}},
  \citenamefont {{Ajello}}, \citenamefont {{Allafort}}, \citenamefont
  {{Baldini}}, \citenamefont {{Ballet}}, \citenamefont {{Barbiellini}},
  \citenamefont {{Baring}}, \citenamefont {{Bastieri}}, \citenamefont
  {{Belfiore}}, \citenamefont {{Bellazzini}},\ and\ \citenamefont
  {et~al.}}]{FermiPulsar2013}%
  \BibitemOpen
  \bibfield  {author} {\bibinfo {author} {\bibfnamefont {A.~A.}\ \bibnamefont
  {{Abdo}}}, \bibinfo {author} {\bibfnamefont {M.}~\bibnamefont {{Ajello}}},
  \bibinfo {author} {\bibfnamefont {A.}~\bibnamefont {{Allafort}}}, \bibinfo
  {author} {\bibfnamefont {L.}~\bibnamefont {{Baldini}}}, \bibinfo {author}
  {\bibfnamefont {J.}~\bibnamefont {{Ballet}}}, \bibinfo {author}
  {\bibfnamefont {G.}~\bibnamefont {{Barbiellini}}}, \bibinfo {author}
  {\bibfnamefont {M.~G.}\ \bibnamefont {{Baring}}}, \bibinfo {author}
  {\bibfnamefont {D.}~\bibnamefont {{Bastieri}}}, \bibinfo {author}
  {\bibfnamefont {A.}~\bibnamefont {{Belfiore}}}, \bibinfo {author}
  {\bibfnamefont {R.}~\bibnamefont {{Bellazzini}}}, \ and\ \bibinfo {author}
  {\bibnamefont {et~al.}},\ }\href {\doibase 10.1088/0067-0049/208/2/17}
  {\bibfield  {journal} {\bibinfo  {journal} {\apjs}\ }\textbf {\bibinfo
  {volume} {208}},\ \bibinfo {eid} {17} (\bibinfo {year} {2013})},\ \Eprint
  {http://arxiv.org/abs/1305.4385} {arXiv:1305.4385 [astro-ph.HE]} \BibitemShut
  {NoStop}%
\bibitem [{\citenamefont {{Cholis}}\ \emph {et~al.}(2014)\citenamefont
  {{Cholis}}, \citenamefont {{Hooper}},\ and\ \citenamefont
  {{Linden}}}]{CholisHooperLinden2014}%
  \BibitemOpen
  \bibfield  {author} {\bibinfo {author} {\bibfnamefont {I.}~\bibnamefont
  {{Cholis}}}, \bibinfo {author} {\bibfnamefont {D.}~\bibnamefont {{Hooper}}},
  \ and\ \bibinfo {author} {\bibfnamefont {T.}~\bibnamefont {{Linden}}},\
  }\href@noop {} {\bibfield  {journal} {\bibinfo  {journal} {ArXiv e-prints}\ }
  (\bibinfo {year} {2014})},\ \Eprint {http://arxiv.org/abs/1407.5583}
  {arXiv:1407.5583 [astro-ph.HE]} \BibitemShut {NoStop}%
\bibitem [{\citenamefont {Carlson}\ and\ \citenamefont
  {Profumo}(2014)}]{CarlsonProfumo}%
  \BibitemOpen
  \bibfield  {author} {\bibinfo {author} {\bibfnamefont {E.}~\bibnamefont
  {Carlson}}\ and\ \bibinfo {author} {\bibfnamefont {S.}~\bibnamefont
  {Profumo}},\ }\href@noop {} {\bibfield  {journal} {\bibinfo  {journal} {ArXiv
  e-prints}\ } (\bibinfo {year} {2014})},\ \Eprint
  {http://arxiv.org/abs/1405.7685} {arXiv:1405.7685 [astro-ph.HE]} \BibitemShut
  {NoStop}%
\bibitem [{\citenamefont {{Petrovi{\'c}}}\ \emph {et~al.}(2014)\citenamefont
  {{Petrovi{\'c}}}, \citenamefont {{Dario Serpico}},\ and\ \citenamefont
  {{Zaharija{\v s}}}}]{Petrovic2014}%
  \BibitemOpen
  \bibfield  {author} {\bibinfo {author} {\bibfnamefont {J.}~\bibnamefont
  {{Petrovi{\'c}}}}, \bibinfo {author} {\bibfnamefont {P.}~\bibnamefont {{Dario
  Serpico}}}, \ and\ \bibinfo {author} {\bibfnamefont {G.}~\bibnamefont
  {{Zaharija{\v s}}}},\ }\href {\doibase 10.1088/1475-7516/2014/10/052}
  {\bibfield  {journal} {\bibinfo  {journal} {JCAP}\ }\textbf {\bibinfo
  {volume} {10}},\ \bibinfo {eid} {052} (\bibinfo {year} {2014})},\ \Eprint
  {http://arxiv.org/abs/1405.7928} {arXiv:1405.7928 [astro-ph.HE]} \BibitemShut
  {NoStop}%
\bibitem [{\citenamefont {Cholis}\ \emph {et~al.}(2015)\citenamefont {Cholis},
  \citenamefont {Evoli}, \citenamefont {Calore}, \citenamefont {Linden},
  \citenamefont {Weniger} \emph {et~al.}}]{Cholis2015}%
  \BibitemOpen
  \bibfield  {author} {\bibinfo {author} {\bibfnamefont {I.}~\bibnamefont
  {Cholis}}, \bibinfo {author} {\bibfnamefont {C.}~\bibnamefont {Evoli}},
  \bibinfo {author} {\bibfnamefont {F.}~\bibnamefont {Calore}}, \bibinfo
  {author} {\bibfnamefont {T.}~\bibnamefont {Linden}}, \bibinfo {author}
  {\bibfnamefont {C.}~\bibnamefont {Weniger}},  \emph {et~al.},\ }\href@noop {}
  {\bibfield  {journal} {\bibinfo  {journal} {ArXiv e-prints}\ } (\bibinfo
  {year} {2015})},\ \Eprint {http://arxiv.org/abs/1506.05119} {arXiv:1506.05119
  [astro-ph.HE]} \BibitemShut {NoStop}%
\bibitem [{\citenamefont {Yang}\ and\ \citenamefont
  {Aharonian}(2016)}]{YangAharonian2016}%
  \BibitemOpen
  \bibfield  {author} {\bibinfo {author} {\bibfnamefont {R.-z.}\ \bibnamefont
  {Yang}}\ and\ \bibinfo {author} {\bibfnamefont {F.}~\bibnamefont
  {Aharonian}},\ }\href@noop {} {\  (\bibinfo {year} {2016})},\ \Eprint
  {http://arxiv.org/abs/1602.06764} {arXiv:1602.06764 [astro-ph.HE]}
  \BibitemShut {NoStop}%
\bibitem [{\citenamefont {Calore}\ \emph {et~al.}(2015)\citenamefont {Calore},
  \citenamefont {Cholis}, \citenamefont {McCabe},\ and\ \citenamefont
  {Weniger}}]{Caloreetal:Taleoftails}%
  \BibitemOpen
  \bibfield  {author} {\bibinfo {author} {\bibfnamefont {F.}~\bibnamefont
  {Calore}}, \bibinfo {author} {\bibfnamefont {I.}~\bibnamefont {Cholis}},
  \bibinfo {author} {\bibfnamefont {C.}~\bibnamefont {McCabe}}, \ and\ \bibinfo
  {author} {\bibfnamefont {C.}~\bibnamefont {Weniger}},\ }\href {\doibase
  10.1103/PhysRevD.91.063003} {\bibfield  {journal} {\bibinfo  {journal} {Phys.
  Rev.}\ }\textbf {\bibinfo {volume} {D91}},\ \bibinfo {pages} {063003}
  (\bibinfo {year} {2015})},\ \Eprint {http://arxiv.org/abs/1411.4647}
  {arXiv:1411.4647 [hep-ph]} \BibitemShut {NoStop}%
\bibitem [{\citenamefont {Abazajian}\ and\ \citenamefont
  {Keeley}(2015)}]{AbazajianKeeley2015}%
  \BibitemOpen
  \bibfield  {author} {\bibinfo {author} {\bibfnamefont {K.~N.}\ \bibnamefont
  {Abazajian}}\ and\ \bibinfo {author} {\bibfnamefont {R.~E.}\ \bibnamefont
  {Keeley}},\ }\href@noop {} {\  (\bibinfo {year} {2015})},\ \Eprint
  {http://arxiv.org/abs/1510.06424} {arXiv:1510.06424 [hep-ph]} \BibitemShut
  {NoStop}%
\bibitem [{\citenamefont {{Geringer-Sameth}}\ \emph
  {et~al.}(2015{\natexlab{a}})\citenamefont {{Geringer-Sameth}}, \citenamefont
  {{Koushiappas}},\ and\ \citenamefont
  {{Walker}}}]{Geringer-SamethKoushiappasWalker2015}%
  \BibitemOpen
  \bibfield  {author} {\bibinfo {author} {\bibfnamefont {A.}~\bibnamefont
  {{Geringer-Sameth}}}, \bibinfo {author} {\bibfnamefont {S.~M.}\ \bibnamefont
  {{Koushiappas}}}, \ and\ \bibinfo {author} {\bibfnamefont {M.~G.}\
  \bibnamefont {{Walker}}},\ }\href {\doibase 10.1103/PhysRevD.91.083535}
  {\bibfield  {journal} {\bibinfo  {journal} {\prd}\ }\textbf {\bibinfo
  {volume} {91}},\ \bibinfo {eid} {083535} (\bibinfo {year}
  {2015}{\natexlab{a}})},\ \Eprint {http://arxiv.org/abs/1410.2242}
  {arXiv:1410.2242} \BibitemShut {NoStop}%
\bibitem [{\citenamefont {{Fermi-LAT
  Collaboration}}(2015)}]{Fermi-LATDwarfSpheroidal2015}%
  \BibitemOpen
  \bibfield  {author} {\bibinfo {author} {\bibnamefont {{Fermi-LAT
  Collaboration}}},\ }\href@noop {} {\bibfield  {journal} {\bibinfo  {journal}
  {ArXiv e-prints}\ } (\bibinfo {year} {2015})},\ \Eprint
  {http://arxiv.org/abs/1503.02641} {arXiv:1503.02641 [astro-ph.HE]}
  \BibitemShut {NoStop}%
\bibitem [{\citenamefont {{Geringer-Sameth}}\ \emph
  {et~al.}(2015{\natexlab{b}})\citenamefont {{Geringer-Sameth}}, \citenamefont
  {{Walker}}, \citenamefont {{Koushiappas}}, \citenamefont {{Koposov}},
  \citenamefont {{Belokurov}}, \citenamefont {{Torrealba}},\ and\ \citenamefont
  {{Evans}}}]{Geringer-Sameth2015}%
  \BibitemOpen
  \bibfield  {author} {\bibinfo {author} {\bibfnamefont {A.}~\bibnamefont
  {{Geringer-Sameth}}}, \bibinfo {author} {\bibfnamefont {M.~G.}\ \bibnamefont
  {{Walker}}}, \bibinfo {author} {\bibfnamefont {S.~M.}\ \bibnamefont
  {{Koushiappas}}}, \bibinfo {author} {\bibfnamefont {S.~E.}\ \bibnamefont
  {{Koposov}}}, \bibinfo {author} {\bibfnamefont {V.}~\bibnamefont
  {{Belokurov}}}, \bibinfo {author} {\bibfnamefont {G.}~\bibnamefont
  {{Torrealba}}}, \ and\ \bibinfo {author} {\bibfnamefont {N.~W.}\ \bibnamefont
  {{Evans}}},\ }\href {\doibase 10.1103/PhysRevLett.115.081101} {\bibfield
  {journal} {\bibinfo  {journal} {Physical Review Letters}\ }\textbf {\bibinfo
  {volume} {115}},\ \bibinfo {eid} {081101} (\bibinfo {year}
  {2015}{\natexlab{b}})},\ \Eprint {http://arxiv.org/abs/1503.02320}
  {arXiv:1503.02320 [astro-ph.HE]} \BibitemShut {NoStop}%
\bibitem [{\citenamefont {{Cirelli}}\ \emph {et~al.}(2013)\citenamefont
  {{Cirelli}}, \citenamefont {{Serpico}},\ and\ \citenamefont
  {{Zaharijas}}}]{CirelliSerpicoZaharijas2013}%
  \BibitemOpen
  \bibfield  {author} {\bibinfo {author} {\bibfnamefont {M.}~\bibnamefont
  {{Cirelli}}}, \bibinfo {author} {\bibfnamefont {P.~D.}\ \bibnamefont
  {{Serpico}}}, \ and\ \bibinfo {author} {\bibfnamefont {G.}~\bibnamefont
  {{Zaharijas}}},\ }\href {\doibase 10.1088/1475-7516/2013/11/035} {\bibfield
  {journal} {\bibinfo  {journal} {JCAP}\ }\textbf {\bibinfo {volume} {11}},\
  \bibinfo {eid} {035} (\bibinfo {year} {2013})},\ \Eprint
  {http://arxiv.org/abs/1307.7152} {arXiv:1307.7152 [astro-ph.HE]} \BibitemShut
  {NoStop}%
\bibitem [{\citenamefont {{Buch}}\ \emph {et~al.}(2015)\citenamefont {{Buch}},
  \citenamefont {{Cirelli}}, \citenamefont {{Giesen}},\ and\ \citenamefont
  {{Taoso}}}]{Cirelli_cookbook_secondaries}%
  \BibitemOpen
  \bibfield  {author} {\bibinfo {author} {\bibfnamefont {J.}~\bibnamefont
  {{Buch}}}, \bibinfo {author} {\bibfnamefont {M.}~\bibnamefont {{Cirelli}}},
  \bibinfo {author} {\bibfnamefont {G.}~\bibnamefont {{Giesen}}}, \ and\
  \bibinfo {author} {\bibfnamefont {M.}~\bibnamefont {{Taoso}}},\ }\href
  {\doibase 10.1088/1475-7516/2015/09/037} {\bibfield  {journal} {\bibinfo
  {journal} {JCAP}\ }\textbf {\bibinfo {volume} {9}},\ \bibinfo {eid} {037}
  (\bibinfo {year} {2015})},\ \Eprint {http://arxiv.org/abs/1505.01049}
  {arXiv:1505.01049 [hep-ph]} \BibitemShut {NoStop}%
\bibitem [{\citenamefont {{G{\'o}mez-Vargas}}\ \emph
  {et~al.}(2013)\citenamefont {{G{\'o}mez-Vargas}}, \citenamefont
  {{S{\'a}nchez-Conde}}, \citenamefont {{Huh}}, \citenamefont {{Peir{\'o}}},
  \citenamefont {{Prada}}, \citenamefont {{Morselli}}, \citenamefont
  {{Klypin}}, \citenamefont {{Cerde{\~n}o}}, \citenamefont {{Mambrini}},\ and\
  \citenamefont {{Mu{\~n}oz}}}]{gomez2013}%
  \BibitemOpen
  \bibfield  {author} {\bibinfo {author} {\bibfnamefont {G.~A.}\ \bibnamefont
  {{G{\'o}mez-Vargas}}}, \bibinfo {author} {\bibfnamefont {M.~A.}\ \bibnamefont
  {{S{\'a}nchez-Conde}}}, \bibinfo {author} {\bibfnamefont {J.-H.}\
  \bibnamefont {{Huh}}}, \bibinfo {author} {\bibfnamefont {M.}~\bibnamefont
  {{Peir{\'o}}}}, \bibinfo {author} {\bibfnamefont {F.}~\bibnamefont
  {{Prada}}}, \bibinfo {author} {\bibfnamefont {A.}~\bibnamefont {{Morselli}}},
  \bibinfo {author} {\bibfnamefont {A.}~\bibnamefont {{Klypin}}}, \bibinfo
  {author} {\bibfnamefont {D.~G.}\ \bibnamefont {{Cerde{\~n}o}}}, \bibinfo
  {author} {\bibfnamefont {Y.}~\bibnamefont {{Mambrini}}}, \ and\ \bibinfo
  {author} {\bibfnamefont {C.}~\bibnamefont {{Mu{\~n}oz}}},\ }\href {\doibase
  10.1088/1475-7516/2013/10/029} {\bibfield  {journal} {\bibinfo  {journal}
  {JCAP}\ }\textbf {\bibinfo {volume} {10}},\ \bibinfo {eid} {029} (\bibinfo
  {year} {2013})},\ \Eprint {http://arxiv.org/abs/1308.3515} {arXiv:1308.3515
  [astro-ph.HE]} \BibitemShut {NoStop}%
\bibitem [{\citenamefont {{Lacroix}}\ \emph {et~al.}(014b)\citenamefont
  {{Lacroix}}, \citenamefont {{B{\oe}hm}},\ and\ \citenamefont
  {{Silk}}}]{Lacroix:2014}%
  \BibitemOpen
  \bibfield  {author} {\bibinfo {author} {\bibfnamefont {T.}~\bibnamefont
  {{Lacroix}}}, \bibinfo {author} {\bibfnamefont {C.}~\bibnamefont
  {{B{\oe}hm}}}, \ and\ \bibinfo {author} {\bibfnamefont {J.}~\bibnamefont
  {{Silk}}},\ }\href {\doibase 10.1103/PhysRevD.90.043508} {\bibfield
  {journal} {\bibinfo  {journal} {Phys. Rev. D}\ }\textbf {\bibinfo {volume}
  {90}},\ \bibinfo {eid} {043508} (\bibinfo {year} {{2014b}})},\ \Eprint
  {http://arxiv.org/abs/1403.1987} {arXiv:1403.1987 [astro-ph.HE]} \BibitemShut
  {NoStop}%
\bibitem [{\citenamefont {{Abazajian}}\ \emph {et~al.}(2015)\citenamefont
  {{Abazajian}}, \citenamefont {{Canac}}, \citenamefont {{Horiuchi}},
  \citenamefont {{Kaplinghat}},\ and\ \citenamefont {{Kwa}}}]{Abazajian2015}%
  \BibitemOpen
  \bibfield  {author} {\bibinfo {author} {\bibfnamefont {K.~N.}\ \bibnamefont
  {{Abazajian}}}, \bibinfo {author} {\bibfnamefont {N.}~\bibnamefont
  {{Canac}}}, \bibinfo {author} {\bibfnamefont {S.}~\bibnamefont {{Horiuchi}}},
  \bibinfo {author} {\bibfnamefont {M.}~\bibnamefont {{Kaplinghat}}}, \ and\
  \bibinfo {author} {\bibfnamefont {A.}~\bibnamefont {{Kwa}}},\ }\href
  {\doibase 10.1088/1475-7516/2015/07/013} {\bibfield  {journal} {\bibinfo
  {journal} {JCAP}\ }\textbf {\bibinfo {volume} {7}},\ \bibinfo {eid} {013}
  (\bibinfo {year} {2015})},\ \Eprint {http://arxiv.org/abs/1410.6168}
  {arXiv:1410.6168 [astro-ph.HE]} \BibitemShut {NoStop}%
\bibitem [{\citenamefont {{Yuan}}\ and\ \citenamefont {{Ioka}}(2015)}]{Ioka}%
  \BibitemOpen
  \bibfield  {author} {\bibinfo {author} {\bibfnamefont {Q.}~\bibnamefont
  {{Yuan}}}\ and\ \bibinfo {author} {\bibfnamefont {K.}~\bibnamefont
  {{Ioka}}},\ }\href {\doibase 10.1088/0004-637X/802/2/124} {\bibfield
  {journal} {\bibinfo  {journal} {\apj}\ }\textbf {\bibinfo {volume} {802}},\
  \bibinfo {eid} {124} (\bibinfo {year} {2015})},\ \Eprint
  {http://arxiv.org/abs/1411.4363} {arXiv:1411.4363 [astro-ph.HE]} \BibitemShut
  {NoStop}%
\bibitem [{\citenamefont {{Kaplinghat}}\ \emph {et~al.}(2015)\citenamefont
  {{Kaplinghat}}, \citenamefont {{Linden}},\ and\ \citenamefont
  {{Yu}}}]{Kaplinghat2015}%
  \BibitemOpen
  \bibfield  {author} {\bibinfo {author} {\bibfnamefont {M.}~\bibnamefont
  {{Kaplinghat}}}, \bibinfo {author} {\bibfnamefont {T.}~\bibnamefont
  {{Linden}}}, \ and\ \bibinfo {author} {\bibfnamefont {H.-B.}\ \bibnamefont
  {{Yu}}},\ }\href {\doibase 10.1103/PhysRevLett.114.211303} {\bibfield
  {journal} {\bibinfo  {journal} {Physical Review Letters}\ }\textbf {\bibinfo
  {volume} {114}},\ \bibinfo {eid} {211303} (\bibinfo {year} {2015})},\ \Eprint
  {http://arxiv.org/abs/1501.03507} {arXiv:1501.03507 [hep-ph]} \BibitemShut
  {NoStop}%
\bibitem [{\citenamefont {{Cirelli}}\ \emph {et~al.}(2011)\citenamefont
  {{Cirelli}}, \citenamefont {{Corcella}}, \citenamefont {{Hektor}},
  \citenamefont {{H{\"u}tsi}}, \citenamefont {{Kadastik}}, \citenamefont
  {{Panci}}, \citenamefont {{Raidal}}, \citenamefont {{Sala}},\ and\
  \citenamefont {{Strumia}}}]{Cirelli_cookbook}%
  \BibitemOpen
  \bibfield  {author} {\bibinfo {author} {\bibfnamefont {M.}~\bibnamefont
  {{Cirelli}}}, \bibinfo {author} {\bibfnamefont {G.}~\bibnamefont
  {{Corcella}}}, \bibinfo {author} {\bibfnamefont {A.}~\bibnamefont
  {{Hektor}}}, \bibinfo {author} {\bibfnamefont {G.}~\bibnamefont
  {{H{\"u}tsi}}}, \bibinfo {author} {\bibfnamefont {M.}~\bibnamefont
  {{Kadastik}}}, \bibinfo {author} {\bibfnamefont {P.}~\bibnamefont {{Panci}}},
  \bibinfo {author} {\bibfnamefont {M.}~\bibnamefont {{Raidal}}}, \bibinfo
  {author} {\bibfnamefont {F.}~\bibnamefont {{Sala}}}, \ and\ \bibinfo {author}
  {\bibfnamefont {A.}~\bibnamefont {{Strumia}}},\ }\href {\doibase
  10.1088/1475-7516/2011/03/051} {\bibfield  {journal} {\bibinfo  {journal}
  {JCAP}\ }\textbf {\bibinfo {volume} {3}},\ \bibinfo {eid} {051} (\bibinfo
  {year} {2011})},\ \Eprint {http://arxiv.org/abs/1012.4515} {arXiv:1012.4515
  [hep-ph]} \BibitemShut {NoStop}%
\bibitem [{\citenamefont {{Cirelli}}\ \emph {et~al.}(2014)\citenamefont
  {{Cirelli}}, \citenamefont {{Gaggero}}, \citenamefont {{Giesen}},
  \citenamefont {{Taoso}},\ and\ \citenamefont
  {{Urbano}}}]{Cirelli_et_al_constraints}%
  \BibitemOpen
  \bibfield  {author} {\bibinfo {author} {\bibfnamefont {M.}~\bibnamefont
  {{Cirelli}}}, \bibinfo {author} {\bibfnamefont {D.}~\bibnamefont
  {{Gaggero}}}, \bibinfo {author} {\bibfnamefont {G.}~\bibnamefont {{Giesen}}},
  \bibinfo {author} {\bibfnamefont {M.}~\bibnamefont {{Taoso}}}, \ and\
  \bibinfo {author} {\bibfnamefont {A.}~\bibnamefont {{Urbano}}},\ }\href
  {\doibase 10.1088/1475-7516/2014/12/045} {\bibfield  {journal} {\bibinfo
  {journal} {JCAP}\ }\textbf {\bibinfo {volume} {12}},\ \bibinfo {eid} {045}
  (\bibinfo {year} {2014})},\ \Eprint {http://arxiv.org/abs/1407.2173}
  {arXiv:1407.2173 [hep-ph]} \BibitemShut {NoStop}%
\bibitem [{\citenamefont {{Evoli}}\ \emph {et~al.}(2008)\citenamefont
  {{Evoli}}, \citenamefont {{Gaggero}}, \citenamefont {{Grasso}},\ and\
  \citenamefont {{Maccione}}}]{Evoli2008}%
  \BibitemOpen
  \bibfield  {author} {\bibinfo {author} {\bibfnamefont {C.}~\bibnamefont
  {{Evoli}}}, \bibinfo {author} {\bibfnamefont {D.}~\bibnamefont {{Gaggero}}},
  \bibinfo {author} {\bibfnamefont {D.}~\bibnamefont {{Grasso}}}, \ and\
  \bibinfo {author} {\bibfnamefont {L.}~\bibnamefont {{Maccione}}},\ }\href
  {\doibase 10.1088/1475-7516/2008/10/018} {\bibfield  {journal} {\bibinfo
  {journal} {JCAP}\ }\textbf {\bibinfo {volume} {10}},\ \bibinfo {eid} {018}
  (\bibinfo {year} {2008})},\ \Eprint {http://arxiv.org/abs/0807.4730}
  {arXiv:0807.4730} \BibitemShut {NoStop}%
\bibitem [{\citenamefont {Cirelli}\ \emph {et~al.}(2014)\citenamefont
  {Cirelli}, \citenamefont {Gaggero}, \citenamefont {Giesen}, \citenamefont
  {Taoso},\ and\ \citenamefont {Urbano}}]{Cirelli:2014lwa}%
  \BibitemOpen
  \bibfield  {author} {\bibinfo {author} {\bibfnamefont {M.}~\bibnamefont
  {Cirelli}}, \bibinfo {author} {\bibfnamefont {D.}~\bibnamefont {Gaggero}},
  \bibinfo {author} {\bibfnamefont {G.}~\bibnamefont {Giesen}}, \bibinfo
  {author} {\bibfnamefont {M.}~\bibnamefont {Taoso}}, \ and\ \bibinfo {author}
  {\bibfnamefont {A.}~\bibnamefont {Urbano}},\ }\href@noop {} {\  (\bibinfo
  {year} {2014})},\ \Eprint {http://arxiv.org/abs/1407.2173} {arXiv:1407.2173
  [hep-ph]} \BibitemShut {NoStop}%
\bibitem [{\citenamefont {{Atwood}}\ \emph {et~al.}(2009)\citenamefont
  {{Atwood}}, \citenamefont {{Abdo}}, \citenamefont {{Ackermann}},
  \citenamefont {{Althouse}}, \citenamefont {{Anderson}}, \citenamefont
  {{Axelsson}}, \citenamefont {{Baldini}}, \citenamefont {{Ballet}},
  \citenamefont {{Band}}, \citenamefont {{Barbiellini}},\ and\ \citenamefont
  {et~al.}}]{2009ApJ...697.1071A}%
  \BibitemOpen
  \bibfield  {author} {\bibinfo {author} {\bibfnamefont {W.~B.}\ \bibnamefont
  {{Atwood}}}, \bibinfo {author} {\bibfnamefont {A.~A.}\ \bibnamefont
  {{Abdo}}}, \bibinfo {author} {\bibfnamefont {M.}~\bibnamefont {{Ackermann}}},
  \bibinfo {author} {\bibfnamefont {W.}~\bibnamefont {{Althouse}}}, \bibinfo
  {author} {\bibfnamefont {B.}~\bibnamefont {{Anderson}}}, \bibinfo {author}
  {\bibfnamefont {M.}~\bibnamefont {{Axelsson}}}, \bibinfo {author}
  {\bibfnamefont {L.}~\bibnamefont {{Baldini}}}, \bibinfo {author}
  {\bibfnamefont {J.}~\bibnamefont {{Ballet}}}, \bibinfo {author}
  {\bibfnamefont {D.~L.}\ \bibnamefont {{Band}}}, \bibinfo {author}
  {\bibfnamefont {G.}~\bibnamefont {{Barbiellini}}}, \ and\ \bibinfo {author}
  {\bibnamefont {et~al.}},\ }\href {\doibase 10.1088/0004-637X/697/2/1071}
  {\bibfield  {journal} {\bibinfo  {journal} {\apj}\ }\textbf {\bibinfo
  {volume} {697}},\ \bibinfo {pages} {1071} (\bibinfo {year} {2009})},\ \Eprint
  {http://arxiv.org/abs/0902.1089} {arXiv:0902.1089 [astro-ph.IM]} \BibitemShut
  {NoStop}%
\bibitem [{\citenamefont {{Acero}}\ \emph {et~al.}(2015)\citenamefont
  {{Acero}}, \citenamefont {{Ackermann}}, \citenamefont {{Ajello}},
  \citenamefont {{Albert}} \emph {et~al.}}]{3FGL}%
  \BibitemOpen
  \bibfield  {author} {\bibinfo {author} {\bibfnamefont {F.}~\bibnamefont
  {{Acero}}}, \bibinfo {author} {\bibfnamefont {M.}~\bibnamefont
  {{Ackermann}}}, \bibinfo {author} {\bibfnamefont {M.}~\bibnamefont
  {{Ajello}}}, \bibinfo {author} {\bibfnamefont {A.}~\bibnamefont {{Albert}}},
  \emph {et~al.} (\bibinfo {collaboration} {Fermi-LAT}),\ }\href {\doibase
  10.1088/0067-0049/218/2/23} {\bibfield  {journal} {\bibinfo  {journal}
  {\apjs}\ }\textbf {\bibinfo {volume} {218}},\ \bibinfo {eid} {23} (\bibinfo
  {year} {2015})},\ \Eprint {http://arxiv.org/abs/1501.02003} {arXiv:1501.02003
  [astro-ph.HE]} \BibitemShut {NoStop}%
\bibitem [{\citenamefont {{Lande}}\ \emph {et~al.}(2012)\citenamefont
  {{Lande}}, \citenamefont {{Ackermann}}, \citenamefont {{Allafort}},
  \citenamefont {{Ballet}}, \citenamefont {{Bechtol}}, \citenamefont
  {{Burnett}}, \citenamefont {{Cohen-Tanugi}}, \citenamefont {{Drlica-Wagner}},
  \citenamefont {{Funk}}, \citenamefont {{Giordano}}, \citenamefont
  {{Grondin}}, \citenamefont {{Kerr}},\ and\ \citenamefont
  {{Lemoine-Goumard}}}]{SpatiallyExtended}%
  \BibitemOpen
  \bibfield  {author} {\bibinfo {author} {\bibfnamefont {J.}~\bibnamefont
  {{Lande}}}, \bibinfo {author} {\bibfnamefont {M.}~\bibnamefont
  {{Ackermann}}}, \bibinfo {author} {\bibfnamefont {A.}~\bibnamefont
  {{Allafort}}}, \bibinfo {author} {\bibfnamefont {J.}~\bibnamefont
  {{Ballet}}}, \bibinfo {author} {\bibfnamefont {K.}~\bibnamefont {{Bechtol}}},
  \bibinfo {author} {\bibfnamefont {T.~H.}\ \bibnamefont {{Burnett}}}, \bibinfo
  {author} {\bibfnamefont {J.}~\bibnamefont {{Cohen-Tanugi}}}, \bibinfo
  {author} {\bibfnamefont {A.}~\bibnamefont {{Drlica-Wagner}}}, \bibinfo
  {author} {\bibfnamefont {S.}~\bibnamefont {{Funk}}}, \bibinfo {author}
  {\bibfnamefont {F.}~\bibnamefont {{Giordano}}}, \bibinfo {author}
  {\bibfnamefont {M.-H.}\ \bibnamefont {{Grondin}}}, \bibinfo {author}
  {\bibfnamefont {M.}~\bibnamefont {{Kerr}}}, \ and\ \bibinfo {author}
  {\bibfnamefont {M.}~\bibnamefont {{Lemoine-Goumard}}},\ }\href {\doibase
  10.1088/0004-637X/756/1/5} {\bibfield  {journal} {\bibinfo  {journal} {\apj}\
  }\textbf {\bibinfo {volume} {756}},\ \bibinfo {eid} {5} (\bibinfo {year}
  {2012})},\ \Eprint {http://arxiv.org/abs/1207.0027} {arXiv:1207.0027
  [astro-ph.HE]} \BibitemShut {NoStop}%
\bibitem [{\citenamefont {Nolan}\ \emph {et~al.}(2012)\citenamefont {Nolan}
  \emph {et~al.}}]{2FGL}%
  \BibitemOpen
  \bibfield  {author} {\bibinfo {author} {\bibfnamefont {P.~L.}\ \bibnamefont
  {Nolan}} \emph {et~al.},\ }\href {\doibase 10.1088/0067-0049/199/2/31}
  {\bibfield  {journal} {\bibinfo  {journal} {Astrophys.J.Suppl.}\ }\textbf
  {\bibinfo {volume} {199}},\ \bibinfo {pages} {31} (\bibinfo {year} {2012})},\
  \Eprint {http://arxiv.org/abs/1108.1435} {arXiv:1108.1435 [astro-ph.HE]}
  \BibitemShut {NoStop}%
\bibitem [{\citenamefont {Bergstrom}\ \emph {et~al.}(2013)\citenamefont
  {Bergstrom}, \citenamefont {Bringmann}, \citenamefont {Cholis}, \citenamefont
  {Hooper},\ and\ \citenamefont {Weniger}}]{Bergstrom:2013jra}%
  \BibitemOpen
  \bibfield  {author} {\bibinfo {author} {\bibfnamefont {L.}~\bibnamefont
  {Bergstrom}}, \bibinfo {author} {\bibfnamefont {T.}~\bibnamefont
  {Bringmann}}, \bibinfo {author} {\bibfnamefont {I.}~\bibnamefont {Cholis}},
  \bibinfo {author} {\bibfnamefont {D.}~\bibnamefont {Hooper}}, \ and\ \bibinfo
  {author} {\bibfnamefont {C.}~\bibnamefont {Weniger}},\ }\href {\doibase
  10.1103/PhysRevLett.111.171101} {\bibfield  {journal} {\bibinfo  {journal}
  {Phys. Rev. Lett.}\ }\textbf {\bibinfo {volume} {111}},\ \bibinfo {pages}
  {171101} (\bibinfo {year} {2013})},\ \Eprint {http://arxiv.org/abs/1306.3983}
  {arXiv:1306.3983 [astro-ph.HE]} \BibitemShut {NoStop}%
\bibitem [{\citenamefont {Ibarra}\ \emph {et~al.}(2014)\citenamefont {Ibarra},
  \citenamefont {Lamperstorfer},\ and\ \citenamefont {Silk}}]{Ibarra:2013zia}%
  \BibitemOpen
  \bibfield  {author} {\bibinfo {author} {\bibfnamefont {A.}~\bibnamefont
  {Ibarra}}, \bibinfo {author} {\bibfnamefont {A.~S.}\ \bibnamefont
  {Lamperstorfer}}, \ and\ \bibinfo {author} {\bibfnamefont {J.}~\bibnamefont
  {Silk}},\ }\href {\doibase 10.1103/PhysRevD.89.063539} {\bibfield  {journal}
  {\bibinfo  {journal} {Phys. Rev.}\ }\textbf {\bibinfo {volume} {D89}},\
  \bibinfo {pages} {063539} (\bibinfo {year} {2014})},\ \Eprint
  {http://arxiv.org/abs/1309.2570} {arXiv:1309.2570 [hep-ph]} \BibitemShut
  {NoStop}%
\bibitem [{\citenamefont {Bringmann}\ \emph {et~al.}(2014)\citenamefont
  {Bringmann}, \citenamefont {Vollmann},\ and\ \citenamefont
  {Weniger}}]{Bringmann:2014lpa}%
  \BibitemOpen
  \bibfield  {author} {\bibinfo {author} {\bibfnamefont {T.}~\bibnamefont
  {Bringmann}}, \bibinfo {author} {\bibfnamefont {M.}~\bibnamefont {Vollmann}},
  \ and\ \bibinfo {author} {\bibfnamefont {C.}~\bibnamefont {Weniger}},\ }\href
  {\doibase 10.1103/PhysRevD.90.123001} {\bibfield  {journal} {\bibinfo
  {journal} {Phys. Rev.}\ }\textbf {\bibinfo {volume} {D90}},\ \bibinfo {pages}
  {123001} (\bibinfo {year} {2014})},\ \Eprint {http://arxiv.org/abs/1406.6027}
  {arXiv:1406.6027 [astro-ph.HE]} \BibitemShut {NoStop}%
\bibitem [{\citenamefont {{Blumenthal}}\ and\ \citenamefont
  {{Gould}}(1970)}]{Blumenthal1970}%
  \BibitemOpen
  \bibfield  {author} {\bibinfo {author} {\bibfnamefont {G.~R.}\ \bibnamefont
  {{Blumenthal}}}\ and\ \bibinfo {author} {\bibfnamefont {R.~J.}\ \bibnamefont
  {{Gould}}},\ }\href {\doibase 10.1103/RevModPhys.42.237} {\bibfield
  {journal} {\bibinfo  {journal} {Reviews of Modern Physics}\ }\textbf
  {\bibinfo {volume} {42}},\ \bibinfo {pages} {237} (\bibinfo {year}
  {1970})}\BibitemShut {NoStop}%
\bibitem [{\citenamefont {{Cirelli}}\ and\ \citenamefont
  {{Panci}}(2009)}]{Cirelli2009}%
  \BibitemOpen
  \bibfield  {author} {\bibinfo {author} {\bibfnamefont {M.}~\bibnamefont
  {{Cirelli}}}\ and\ \bibinfo {author} {\bibfnamefont {P.}~\bibnamefont
  {{Panci}}},\ }\href {\doibase 10.1016/j.nuclphysb.2009.06.034} {\bibfield
  {journal} {\bibinfo  {journal} {Nuclear Physics B}\ }\textbf {\bibinfo
  {volume} {821}},\ \bibinfo {pages} {399} (\bibinfo {year} {2009})},\ \Eprint
  {http://arxiv.org/abs/0904.3830} {arXiv:0904.3830 [astro-ph.CO]} \BibitemShut
  {NoStop}%
\end{thebibliography}%

\end{document}